\documentclass[fleqn,10pt]{article}
\pdfoutput=1
\usepackage[utf8]{inputenc}
\usepackage[T1]{fontenc}
\usepackage{graphicx}
\usepackage{siunitx}
\usepackage{hyperref}
\DeclareSIUnit\angstrom{\text {Å}}
\DeclareSIUnit\bar{bar}
\DeclareSIUnit\elementarycharge{\text{\ensuremath{e}}}
\DeclareSIUnit\Molar{\textsc{m}}
\usepackage[superscript]{cite}
\usepackage{authblk}
\usepackage{color,soulutf8}
\usepackage{lineno}
\usepackage{verbatim}
\title{
Detecting dynamic domains and local fluctuations in complex molecular systems \textit{via} timelapse neighbors shuffling}

\author[1]{Martina Crippa}
\author[2]{Annalisa Cardellini} 
\author[1]{Cristina Caruso}
\author[1,2,*]{Giovanni M. Pavan}
\affil[*]{corresponding author: Giovanni M. Pavan (giovanni.pavan@polito.it)}
\affil[1]{Department of Applied Science and Technology, Politecnico di Torino, Corso Duca degli Abruzzi 24, 10129 Torino, Italy}
\affil[2]{Department of Innovative Technologies, University of Applied Sciences and Arts of Southern Switzerland, Polo Universitario Lugano, Campus Est, Via la Santa 1, 6962 Lugano-Viganello, Switzerland}
\date{}

\begin{document}
\maketitle
\begin{abstract}
Many complex molecular systems owe their properties to local dynamic rearrangements or fluctuations that, despite the rise of machine learning (ML) and sophisticated structural descriptors, remain often difficult to detect. Here we show an ML framework based on a new descriptor, named Local Environments and Neighbors Shuffling (LENS), which allows identifying dynamic domains and detecting local fluctuations in a variety of systems \textit{via} tracking how much the surrounding of each molecular unit changes over time in terms of neighbor individuals. Statistical analysis of the LENS time-series data allows to blindly detect different dynamic domains within various types of molecular systems with, \textit{e.g.}, liquid-like, solid-like, or diverse dynamics, and to track local fluctuations emerging within them in an efficient way. The approach is found robust, versatile, and, given the abstract definition of the LENS descriptor, capable of shedding light on the dynamic complexity of a variety of (not necessarily molecular) systems.
\end{abstract}

\section*{Introduction}
Supramolecular assemblies and quasi-crystalline structures, are characterized by a non-trivial internal dynamics that is often ambiguous and challenging to reconstruct.\cite{cho2021,kaser2022,bae2022,baletto2019,cioni2022}
Self-assembled structures, composed of molecular units interacting with each other \textit{via} reversible non-covalent interactions, offer a notable example of systems where a continuous reshuffling and exchange of the constitutive building-blocks is at the origin of interesting bioinspired and stimuli-responsive properties.\cite{bochicchio2017natcomm,gasparotto2020,aida2012,webber2016,savyasachi2017,brunsveld2001,boekhoven2014,lionello2021} 
Also other completely different systems, such as, \textit{e.g.},
metallic structures, are known to possess a non-trivial internal dynamics. Already at $\sim 1/3$ of the melting temperature (\textit{i.e.}, the so-called Hüttig temperature) metal surfaces are known to enter a dynamic equilibrium where atoms may leave their lattice positions and start moving on the atomic surface, inducing surface transformations and reconstructions.\cite{spencer1986, jayanthi1985,cioni2022} In nanosized metal systems (metal nanoclusters, nanoparticles, etc.), such atomic dynamics emerges even at lower (\textit{e.g.}, room) temperature.\cite{rapetti2022}
In all these cases, the dynamics and fluctuations in time of the building blocks are deeply connected to important properties of the materials, such as, \textit{e.g.}, the mechanical properties of metals,\cite{yamakov2004deformation,zepeda2017probing,wang2021atomistic} their performance in heterogeneous catalysis,\cite{koch1992,wang1991,antczak2010,gazzarrini2021} or, for example, the dynamics adaptivity and stimuli-responsiveness of supramolecular materials.\cite{demarco2021,crippa2022,torchi2018,bochicchio2019defects,lionello2021} Gaining the ability to track the dynamics of the building blocks in complex self-organizing molecular systems is fundamental to studying and rationalizing most of their properties.\cite{albertazzi2014,wang2022,sosso2018,bochicchio2017natcomm,bochicchio2019defects,sharp2018} However, this is also typically challenging and demands efficient analysis approaches.

Molecular dynamics (MD) simulations are being increasingly used to obtain high-resolution insights into the behavior of a variety of systems.\cite{bochicchio2018mol}
\cite{frederix2018,Lee2012,Bejagam2015,cho2021,perego2021,bochicchio2017acs,behler2007,bartok2010,behler2016} 
One key advantage of MD trajectories is that these keep track of the motion of the individual molecular units and retain a large amount of high-resolution data useful to reconstruct the complex structural dynamics of the system. Nonetheless, non-trivial aspects concern the extraction of relevant information from the large amount of data contained in the MD trajectories and their conversion to human-readable form. 
Typical descriptors used to extract information from MD trajectories may be divided into system-specific or abstract (general) descriptors. Extensively used to investigate, \textit{e.g.}, ice-water systems,\cite{errington2001} or metal clusters,\cite{behler2007,rossi2018} to cite a few examples, \textit{ad hoc} descriptors build on considerable \textit{a priori} knowledge of the system under consideration and are developed and optimized on such specific system, but poorly transferable to different ones.
Abstract descriptors \textit{e.g.}, Smooth Overlap of Atomic Positions (SOAP), radial distribution functions (\textit{g(r)}), etc. are conversely less specific and more general.\cite{bartok2013,errington2001,behler2011,drautz2019,faber2015,gasparotto2018,musil2021,pietrucci2015} Although less precise than the tailored ones, abstract descriptors offer an advantage in terms of transferability: they can be applied to different systems and do not require deep \textit{a priori} knowledge of the system's features.\cite{gardin2022,bartok2013,musil2021} 
The high-dimensional data obtained using such descriptors are typically converted into lower-dimensional human-readable information \textit{via} supervised and unsupervised machine learning (ML) approaches (\textit{e.g.}, clustering), and analyzed to reconstruct the internal dynamics of the studied systems.\cite{andrews2022,glielmo2021,gasparotto2014,bartok2017,chen2019,davies2022,noe2019} 
For example, unsupervised clustering of SOAP\cite{bartok2013} data extracted from MD trajectories recently allowed to study the complex dynamics in self-assembling fibers, micelles, lipid bilayers, \cite{gasparotto2018,capelli2021,gardin2022,lionello2022,cardellini2022} in confined ionic environments,\cite{gasparotto2018,lionello2022} as well as in metal nanoparticles and surfaces.\cite{cioni2022,rapetti2022} 

Despite the advantages granted by such ML developments, the behavior of complex molecular systems is often determined by rare fluctuations and local dynamic rearrangements,\cite{bochicchio2019defects,bochicchio2017natcomm,gasparotto2020} poorly captured by average-based measurements. The dynamics of defects in materials science is a typical example of local events determining a variety of hierarchical materials' properties.\cite{schaedel2019,sharp2018}
However, detecting and tracking local fluctuations becomes increasingly difficult when dealing with complex molecular/atomic systems where a certain degree of structural order is coupled with a continuous exchange and reshuffling of molecules/atoms.\cite{crippa2022}  
Abstract descriptors that are transferable and at the same time effective in capturing local fluctuations in complex dynamic systems would be fundamental. 

Here we develop an abstract descriptor named "Local Environments and Neighbors Shuffling (LENS)". Combined with a ML-based analysis, LENS is capable of detecting different dynamic domains and tracking local fluctuations in complex molecular systems without deep prior knowledge of the chemical/physical features of the constituent building blocks but simply by tracing their reciprocal motion and instantaneous fluctuations in space and time. LENS builds on a relatively simple definition and can be transferred to a variety of complex systems with,  liquid, solid, or diverse/hybrid dynamics (\textit{e.g.}, typical of phase-transitions). The results obtained with LENS change the vision of complex molecular systems and, building on simple and general basic concepts, suggest a broad applicability (\textit{e.g.}, not necessarily restricted to molecular ones). 

\section*{Results}
\subsection*{LENS: Local Environments \& Neighbors Shuffling}
In this work, we analyze molecular dynamics (MD) trajectories of various molecular/atomic systems, from soft to quasi-crystalline ones, possessing liquid-like to solid-like dynamics. 
As examples of fluid-like systems, we use lipid bilayers and surfactant micelles, \cite{cardellini2022} while for solid-like dynamics, we focus on metal surfaces\cite{cioni2022} and nanoparticles.\cite{rapetti2022}
Furthermore, we also include systems with intrinsically non-uniform internal dynamics, such as, \textit{e.g.}, a system where ice and liquid water coexist in dynamic equilibrium in correspondence of the solid-liquid transition, and soft self-assembled fibers whose behavior is dominated by local dynamic defects (see Supplementary Table 1 for system details).\cite{bochicchio2017natcomm,gasparotto2020,gardin2022}
Such a large diversity is functional to test the generality of our approach.

\begin{figure}
\centering
\includegraphics[width=\columnwidth]{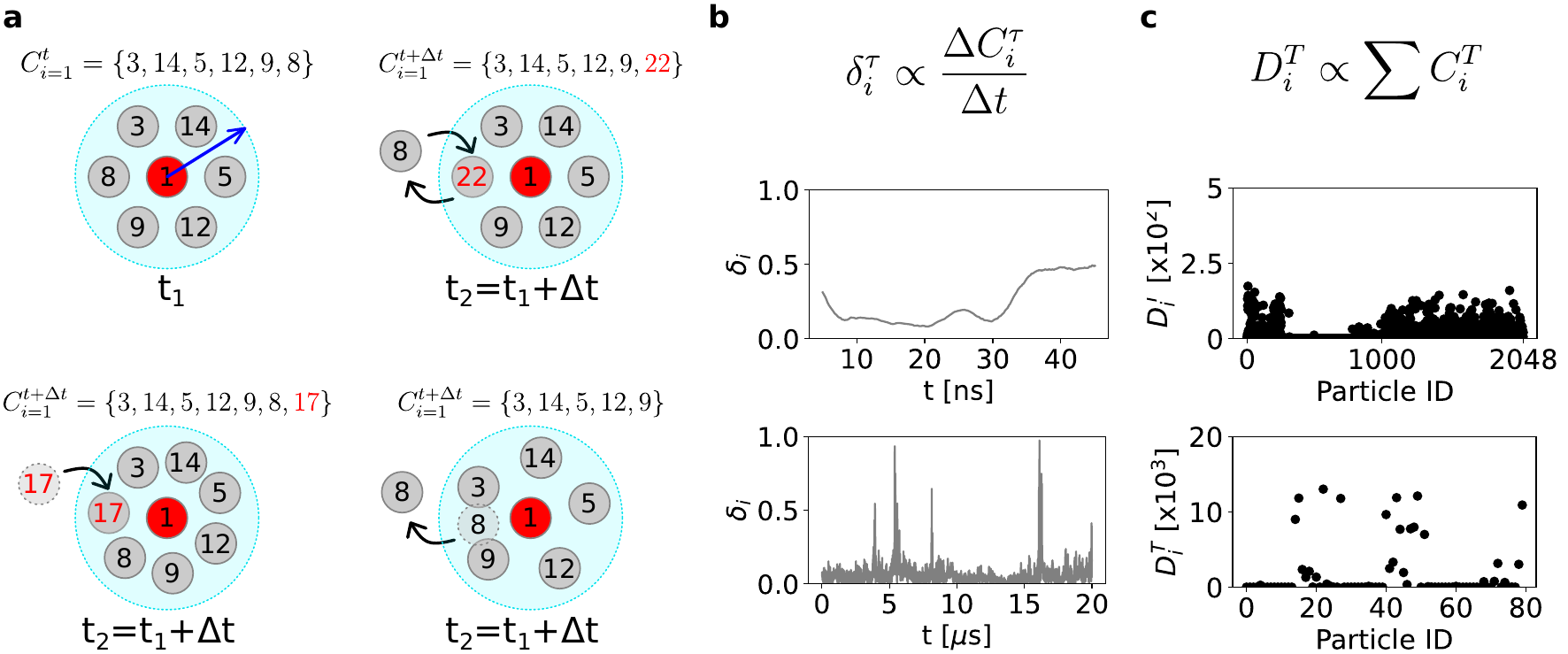}
\caption{Tracking local neighbor environments in complex molecular systems with the LENS descriptor.
(a) The local molecular environment of the particle $i=1$ at time $t$ is defined by an array $C_i^t$ containing the identities (IDs) of all molecular units within a sphere of radius $r_{cut}$ (blue arrow). Along the MD trajectory, $C_i^t$ can be calculated for all constitutive particles at each sampled MD timestep $t$. The local molecular environment $C_i^t$ of the unit $i=1$ (red particle) at time $t_1$ (top-left). The local environment $C_i^{t+\Delta t}$ at time $t_2 = t_1+\Delta t$, when one particle permutation occurs in $\Delta t$ (top-right). The local environment $C_i^{t+\Delta t}$ at time $t_2 = t_1+\Delta t$, when one particle enters (bottom-left) or leaves (bottom-right) the neighborhood sphere in $\Delta t$. (b) The LENS descriptor. The LENS signal for the generic particle $i$ $\delta_i^{\tau}$ is proportional to the number of changes in the neighborhood within a timestep $\tau$ (top). Two examples of typical LENS signals, $\delta_i(t)$ (raw data smoothed as described in the Methods section), for a particle with fluid-like behavior (center) and a particle with dynamics dominated by local fluctuations (bottom). (c) Global statistical analysis. The contact events between the particle $i$ and all the others along a trajectory of T frames are collected and counted in the $D_i^T$ array. Two examples of contact counts, $D_i^T$, between a molecule $i$ and all other IDs in the two distinct dynamics cases of panel (c).}
\label{fig:fig01}
\end{figure}

Despite their intrinsic differences, all these systems can be considered from an abstract point of view as composed of \textit{N} dynamically interacting particles with their own individual trajectories. 
The analysis approach we present herein is based on the concept of molecular individuals (even in cases of systems of chemically identical particles). 
In particular, from the global trajectory of the system, we can identify the sub-trajectory of the \textit{i}th particle (with \textit{i} ranging from 1 to \textit{N}). From this, we can thus describe the local environment surrounding each \textit{i}th particle in terms of its neighbor individuals (IDs) and monitor the changes of IDs at each interval between the sampled timestep $\Delta t$ along the trajectory. Figure \ref{fig:fig01}a (top-left) shows a representative scheme where, at a given time $t$, the neighbor ID units (gray circles) surrounding the \textit{i}th particle ($i=1$ -- red circle) within a sphere of radius $r_{cut}$ (namely, the neighborhood cutoff) are listed in a fingerprint string $C_{i=1}^t$. 
The local $C_{i=1}^{t+\Delta t}$ environment at $t+\Delta t$ may change from that one at time $t$ ($C_{i=1}^t$) when neighbor switching (Figure \ref{fig:fig01}a: top-right -- permutation), addition (Figure \ref{fig:fig01}a: bottom-left), or subtraction (bottom-right) occur in $\Delta t$.

Our analysis is based on monitoring the time-lapse sequence of the ID data along a given trajectory. We developed a new descriptor named "Local Environments and Neighbors Shuffling (LENS)", which allows us to track to what extent the $i$th local environment changes at every consecutive time interval ($C_i^t$, $C_i^{t + \Delta t}$, $C_i^{t + 2 \Delta t}$, etc.) along its trajectory. LENS is built to detect essentially two types of changes in the local neighbor environments along a trajectory: (i) changes in the number of neighbors (addition/leave of one or more neighbors), and/or (ii) changes in the IDs of the neighbors (permutation of one or more neighbor IDs).
The instantaneous value of LENS ($\delta_i$, in its variable form) is defined as: 

\begin{equation}
\centering
    \delta_i^{t+\Delta t}=\frac{\#(C_i^{t} \bigcup C_i^{t+\Delta t} - C_i^{t} \bigcap C_i^{t+\Delta t})}{\#(C_i^t+C_i^{t+\Delta t})}
\label{eq:01}
\end{equation} 

where the first ($C_i^{t} \bigcup C_i^{t+\Delta t}$) and the second term ($C_i^{t} \bigcap C_i^{t+\Delta t}$) of the numerator are respectively the mathematical union and intersection of the neighbor IDs present within $r_{cut}$ from particle $i$ at time $t$ and at time $t + \Delta t$. The denominator contains a normalization factor, which is the total length of the neighbor ID lists (strings) at the two consecutive timesteps.
Thus, for every particle $i$, the $\delta_i(t)$ ranges from 0 to 1 for local neighbor environments which are respectively persistent to highly dynamic over time. 
For example, in the hypothetical case where no local neighbor changes occur in $\Delta t$, the union of $C_i^t$ and $C_i^{t+\Delta t}$ is identical to their intersection, and LENS gives $\delta_i^{t+\Delta t}=0$. In a case where, \textit{e.g.}, all IDs permute in different IDs in $\Delta t$ (complete shuffling while the number of neighbors remains constant), the numerator of the $\delta_i^t$ ($(C_i^t + C_i^{t+\Delta t})-0$) is equal to the denominator, and LENS gives $\delta_i^{t+\Delta t}=1$. As shown in Figure \ref{fig:fig01}b (top), the LENS signal ($\delta_i$) for the generic particle $i$ can be considered proportional to the local neighborhood changes within a time-interval $\Delta t$. Figure \ref{fig:fig01}b reports two examples of LENS signal over time in the cases of a particle with fluid-like behavior (center) and of another particle (bottom) which dynamics is dominated by local fluctuations. 

The time-lapse analysis provided by LENS can be also corroborated/compared with a time-independent statistical analysis of the ID neighbor list data $C_i$. 
In particular, from the ID neighbor list data $C_i$ calculated at every sampled time-step ($t$, $t+\Delta t$, $t+2\Delta t$, etc.), one can easily estimate how many times a particle $i$ has been in direct contact with all the other $N$ ID particles during a sampled trajectory T. 
The inter-IDs contact counts are then stored into an array $D_i^{T}$ (Figure \ref{fig:fig01}c).
In such global statistical analysis, the $D_i^{T}$ data are useful to detect the presence of domains differing from each other in terms of dinamicity/persistence of the local neighbor individuals over time (\textit{i.e.}, in terms of how quickly/slowly the neighbor IDs change along the trajectory). 
In particular, analysis of the global $D^{T}$ contact matrix (Figure \ref{fig:fig02}e) provides information on the propensity of a certain $i$ unit to be, \textit{e.g.}, persistently surrounded by the same neighbors (IDs) or by a population that is in continuous reshuffling during the simulation (see Methods for details). 
As it will be discussed in the next sections, such global time-independent analysis does correlate with the LENS one for systems composed of statistically-relevant dynamically-diverse domains (populated by a relevant number of units that can be effectively detected \textit{via} "dynamic-pattern recognition" approaches), while it does not for systems whose dynamics is dominated by sparse local fluctuations/transitions.

\subsection*{A LENS onto the dynamics of fluid-like systems}

\begin{figure*}[ht!]
\centering
\includegraphics[width=\columnwidth]{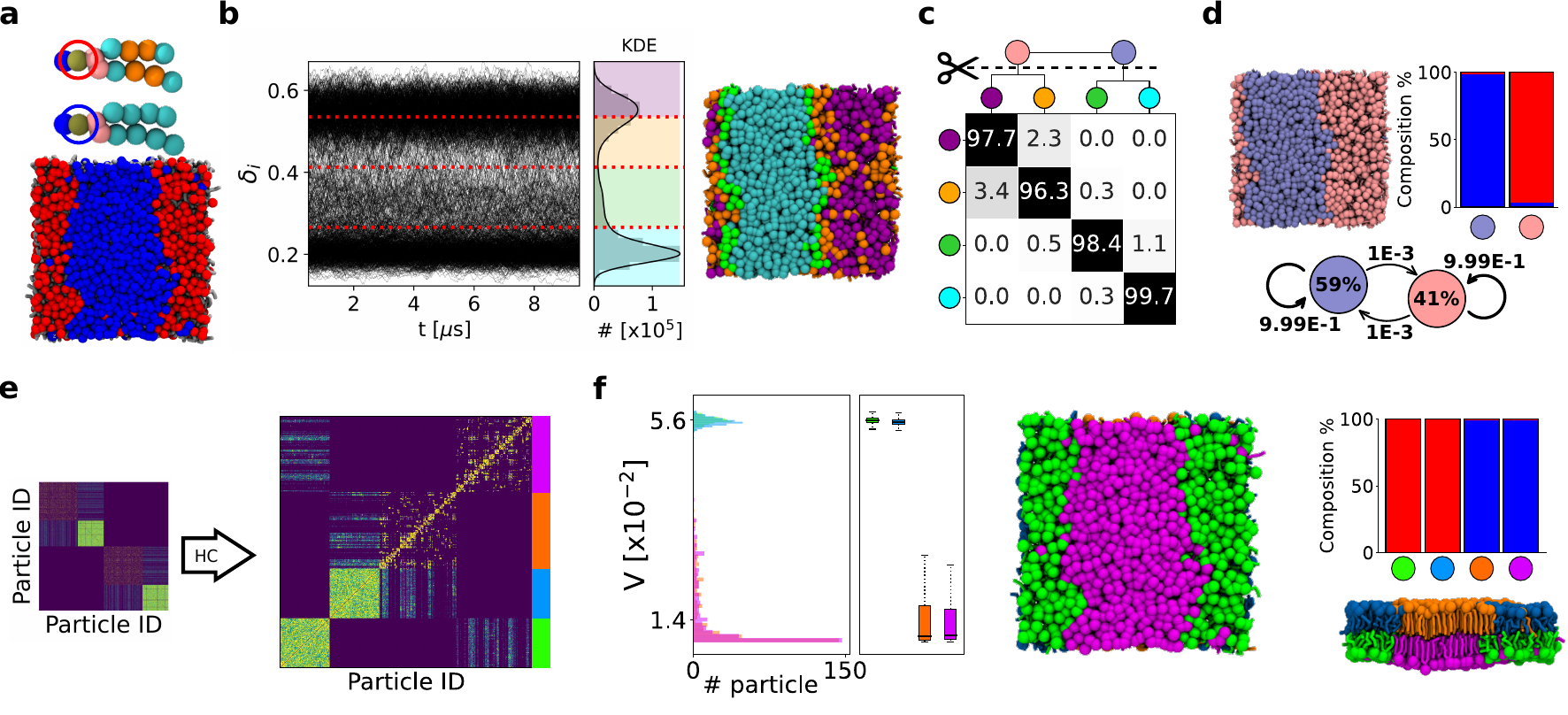}
\caption{LENS analysis of fluid-like systems. (a) Bicomponent lipid bilayer made of $1150$ lipid molecules, namely \textbf{DIPC}:\textbf{DPPC} in $2$:$3$ ratio ($460$:$690$ in total, $230$:$345$ per leaflet) colored in red and blue, respectively. (b) Time-series of LENS signals, $\delta_i(t)$, with the Kernel Density Estimate (KDE) of LENS distribution classified into four clusters (left). MD snapshot of lipids bilayer colored according to their clusters of belonging (right). (c) Inter-clusters normalized transition probability matrix. The $p_{ii}$ and $p_{ij}$ matrix entries indicate the $\%$ probability that molecules with LENS signal typical of a cluster $i$ remain in that dynamical environment or move to another one $j$ (with different dynamics) in $\Delta t$. Hierarchical grouping of the  dynamically-closer clusters (dendrogram cutting) is reported on top of the matrix, and it provides two macroclusters, merging cyan and green on one hand, and orange and purple on the other hand. (d) MD snapshot of lipid bilayer colored according to macroclusters in (c): light-blue identifying \textbf{DPPC} lipids, pink identifying \textbf{DIPC} lipids (top-left). Cluster composition histogram (top-right) and interconversion diagram (bottom) with the transition exchange probabilities and the cluster population percentages (within colored circle). (e) HC analysis of the $D^{T}$ matrix identifying four main clusters (light blue, green, purple, orange). (f) \emph{Variability}, $V$, analysis of the clusters: distributions, median (first and third quartiles), maximum and minimum values (whiskers). The green and light blue clusters, arranging on separated bilayer leaflets, have higher $V$ than the orange and magenta clusters (left). MD snapshot front view of lipid bilayer colored according the HC clustering of $D^{T}$ matrix (middle). Cluster composition histogram (top-right): the green and light blue clusters are made of \textbf{DIPC} lipids (in red in (a)), while the orange and magenta ones correspond to the \textbf{DPPC} lipids (in blue in (a)). MD snapshot lateral view of lipid bilayer colored according the HC clustering of $D^{T}$ matrix (bottom-right).}
\label{fig:fig02}
\end{figure*}

We start testing LENS on a soft molecular system with non-trivial fluid-like dynamics (Figure \ref{fig:fig02}). 
In particular, we analyze a MD simulation trajectory of a coarse-grained (CG) bicomponent lipid bilayer composed of 1150 \textbf{DIPC}:\textbf{DPPC} lipid molecules in $2$:$3$ ratio (see Figure \ref{fig:fig02}a, where \textbf{DIPC} and \textbf{DPPC} are colored in red and blue respectively). 
It is well known that at $T = 280$ \si{\kelvin}, a $2$:$3$ \textbf{DIPC}:\textbf{DPPC} lipid bilayer self-segregates into two distinct regions, populated by the two lipid species which do not mix in such conditions.\cite{baoukina2017} 
For this lipid model we ran $15$ \si{\micro\second} of CG-MD simulation using the Martini 2.2 force field,\cite{marrink2007} (see Methods section and Supplementary Table 1 for details). The last $10$ \si{\micro\second}, representative of an equilibrated MD regime, are used for the analysis. 

Being interested in the lipid shuffling dynamics, in our LENS analysis we use the lipid heads as reference constituent particles and we set a time-interval of $\Delta t = 10$ \si{\nano\second} with a neighborhood cutoff $r_{cut} = 16$ \si{\angstrom} (Supplementary Figure $2$). 
On average, with such a setup, every reference lipid has $\sim 13$ neighbors. Noteworthy, the robustness of the analysis while changing the $r_{cut}$ or $\Delta t$ is demonstrated in Supplementary Figures $3$,$4$. Figure \ref{fig:fig02}b shows on the left the time-profiles of $\delta_i(t)$ for the $1150$ lipid heads forming the bilayer, while on the right the $\delta_i$ data distribution and the correlated KDE are reported. Here, two peaks are clearly detected. A simple supervised clustering analysis, carried out with the KMeans algorithm\cite{lloyd1982} on LENS signals, demonstrates that the $\delta_i$ distribution can be classified into four clusters (cyan, green, orange, and purple) denoted as dynamical clusters or domains. The time-series data of the individual lipid IDs along the trajectory allows computing the exchange probability matrix represented in Figure \ref{fig:fig02}c and obtaining the associated dendrogram detailing the hierarchical interconnection/adjacency between such four detected clusters. In the exchange probability matrix, the $p_{nn}$ and $p_{nm}$ entries indicate the $\%$ probability for a lipid $i$ belonging to a given dynamical cluster $n$ -- having a characteristic rate-of-change of its local neighbor environment -- to remain in that dynamic domain or to undergo a transition into a different dynamic cluster $m$ -- with a different LENS fingerprint -- in $\Delta t$ (see Methods for additional details).
The four obtained microclusters can be then hierarchically merged based on the dendrogram in Figure \ref{fig:fig02}c, by connecting those having a high probability of exchanging molecules. Such approach provides two main macroclusters, colored in light blue and pink, whose populations and transition probabilities in $\Delta t$ are reported in the interconversion diagram of Figure \ref{fig:fig02}d within circles and on the arrows, respectively. 

The data show that the pink domain, obtained after merging orange and purple clusters, is dominated by those lipid units having a higher aptitude to mutate their neighborhood environment: in other words, by those having a more dynamic local neighbor environment (high $\delta_i$). On the other hand, the lipids belonging to the light blue domain, resulting from blending the cyan and green microclusters, reveal a slower variation of their surrounding environment and hence weaker local mobility (low $\delta_i$).  Not surprisingly, while the pink dynamics domain overlaps with the \textbf{DIPC} molecules (red component), known to be in liquid phase \cite{baoukina2017}, the light blue cluster matches up with the \textbf{{DPPC}} lipids (blue component) that are instead in gel phase \cite{baoukina2017} (see composition histogram in Figure \ref{fig:fig02}d, top-right). Furthermore, the estimated exchange probabilities between the pink and light blue macroclusters are very low ($<1\%$) in $\Delta t = 10$ \si{\nano\second}, which is consistent with a sharp segregation between the gel and fluid phases.

Figure \ref{fig:fig02}e,f illustrates the main outcomes of the global statistical analysis explained in the previous paragraph. The collected data, $D_i^{T}$, are organized into a count matrix where the single entry $i$,$j$ defines the total number of neighboring events between lipids $i$ and $j$ (Figure \ref{fig:fig02}e). 
Although such statistical analysis is unrelated to the temporal sequence of the $C_i^t$s, the global $D^{T}$ matrix allows distinguishing the propensity of a certain lipid to be, \textit{e.g.}, persistently surrounded by the same neighbors or by a population in continuous exchange (reshuffling) during the simulation. After hierarchical clustering (HC) of the $D^{T}$ matrix data (see Methods for details), four main dynamic domains are identified (Figure \ref{fig:fig02}e, right): in green, light-blue, orange, purple. Lipid molecules characterized by a similar distribution of neighbor contacts in the $D^{T}$ matrix are classified in the same dynamics domain. For a more quantitative investigation, we also define a \emph{Variability} (\emph{V)} parameter by estimating the standard deviation of the $D_i^{T}$: the broader is the distribution of the neighbor IDs, the higher is the \emph{Variability} (see Methods for details). The analysis shows that the green and light-blue domains are identically highly dynamic, while the orange and purple clusters, similar to each other from a dynamical standpoint, are $\sim4$ times more static (Figure \ref{fig:fig02}f, left). 
Note that, while having the same variability and local-shuffling dynamics, the two green/light-blue (and orange/purple) clusters are identified in this analysis as separate environments. In fact, since the bilayer model replicated on the \textit{xy} through periodic boundary conditions, the \textbf{DIPC} and \textbf{DPPC} lipids belonging to the upper leaflet do not get in contact with those in the bottom one (their $D_i^{T}$ distributions do not overlap). 
The histograms in Figure \ref{fig:fig02}f (right) reveal that the green and blue clusters correspond to red \textbf{DIPC} lipids, while the orange and purple domains correspond to the blue \textbf{DPPC} molecules. This is consistent with the experimental evidence,\cite{baoukina2017} showing that the \textbf{DIPC} lipids form a liquid phase segregating from gel-phase \textbf{DPPC} molecules at the simulation temperature. 
It is worth noting how the macroclusters obtained with the global statistical analysis (Figure \ref{fig:fig02}e,f) correspond in these case to those obtained \textit{via} LENS-based clustering. As anticipated, such correspondence occurs only in those systems composed of "statistically dominant" different dynamic domains, as in this case, where a liquid and a fluid phase coexist in the bilayer system. 
In the next sections, we will also discuss cases where LENS detects fluctuations that get lost and cannot be tracked \textit{via} such global/average analyses, since they are not statistically relevant.

To test the generality of our approach, we also tested the same analysis on a CG-MD simulation trajectory of a bi-component micelle model (Supplementary Figure $2$) made of n-stearoyl L-histidine (\textbf{H}) and p-nitrophenyl ester of n-stearoyl L-phenylalanine self-assembling surfactant molecules (see Methods for details).\cite{cardellini2022} Supplementary Figures $2$a-d show how both LENS and the corresponding time-independent \textit{Variability} analyses identify two distinct dynamic domains: a "donut-like" region of \textbf{H} surfactants (red) and two separated, flatter circular sections of \textbf{F-NP} surfactants (in blue). Similarly to the bi-component lipid bilayer case discussed above, the dynamics of such bi-component micellar assembly appears being thus characterized by different statistically-relevant dynamic domains.

\subsection*{Into phase transitions \& dynamic phases coexistence}

We also tested the efficiency of LENS in characterizing phase transitions as well as the dynamic coexistence between different phases. 
To this end, we discuss two different example systems: (i) a (soft) \textbf{DPPC} lipid bilayer system undergoing gel-to-liquid transition with increasing temperature, and (ii) a simulation box where crystalline ice and liquid water coexist in correspondence of the melting/solidification temperature. 

\begin{figure*}[ht!]
\centering
\includegraphics[width=\columnwidth]{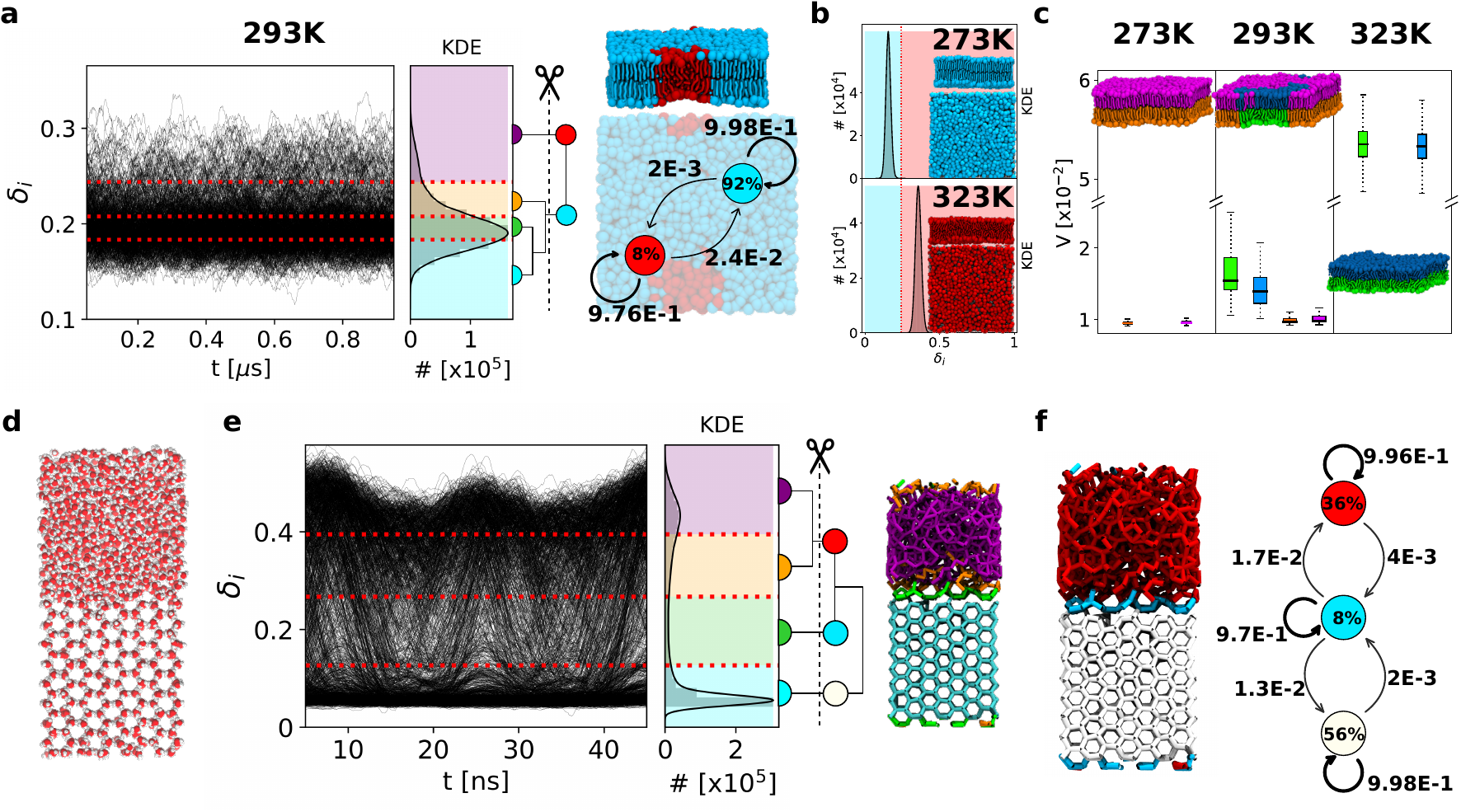}
\caption{LENS analysis of multi-phases coexistence. (a) LENS analysis for \textbf{DPPC} lipid bilayer in coexistence conditions at $T = 293$ \si{\kelvin}: time-series of LENS signals, $\delta_i(t)$, with the KDE of LENS distribution, and the interconnection dendrogram identifying two macroclusters in cyan and red (left). MD snapshot of a \textbf{DPPC} lipid bilayer colored  according to the two main LENS macroclusters (top-right) and related dynamic interconversion diagram (bottom-right). (b) LENS analysis, detecting phase transition at $T = 273$ \si{\kelvin} (gel) and $T = 323$ \si{\kelvin} (liquid) for a \textbf{DPPC} lipid bilayer. (c) Global statistical neighborhood analysis of the \textbf{DPPC} lipid bilayer across a phase transition: at $T = 273$ \si{\kelvin} the bilayer is in gel-state (low variability $V$), at $T = 323$ \si{\kelvin} it is in the liquid-state (high), while two domains (gel and fluid) are detected at $T = 293$ \si{\kelvin}. (d) \textbf{Ice/water} coexistence in an MD simulation (using the TIP4P/Ice water model at $268$ \si{\kelvin}\cite{garcia2006melting}): Oxygen atoms in red and Hydrogen atoms in white. (e) LENS analysis of ice-water coexistence: time-series of LENS signals ($\delta_i(t)$: left) with the KDE of the LENS distribution, and the HC interconnection dendrogram-based clustering (center). The four initially-detected LENS microclusters, represented in different colors in the MD snapshot (right), are merged \textit{via} HC into three main dynamic environments/clusters. (f) Left: MD snapshot showing the three main LENS macroclusters, which identify the liquid phase (in red), the ice phase (in white), and the ice-liquid interface region (in cyan). Right: dynamic interconversion diagram showing how water molecules undergo dynamic transitions from ice-to-liquid and \textit{vice versa}, passing through the ice-liquid interface in such conditions.}
\label{fig:fig03}
\end{figure*}

For case (i), we analyze $1001$ consecutive snapshots taken along 1 \si{\micro\second} of CG-MD simulations ($\Delta t=1$ \si{\nano\second}) of a lipid bilayer model composed of $1152$ self-assembled \textbf{DPPC} lipids parametrized with the Martini force field \cite{marrink2007} at three distinct temperatures: $273$ \si{\kelvin}, $293$ \si{\kelvin}, and $323$ \si{\kelvin} (see Methods for details).\cite{capelli2021} 
It is known that \textbf{DPPC} lipid bilayers have a transition temperature gel-to-liquid of $\sim 315$ \si{\kelvin}.\cite{biltonen1993use} 
However, detecting in a robust manner such gel-liquid phases is not straightforward and typically requires sophisticated analysis approaches that are not always trivial to handle.\cite{capelli2021, baoukina2012}
After reducing the number of clusters detected by KMeans (Supplementary Figure $6$), LENS identifies two main phases dominating the \textbf{DPPC} bilayer at  $T = 293$ \si{\kelvin} (Figure \ref{fig:fig03}a): the $\delta_i(t)$ data indicates that while the largest part of lipids show a reduced local reshuffling of neighbors over time, a non-negligible portion of them is more dynamic.
As shown in Figure \ref{fig:fig03}a (right), two phases coexist at $T = 293$ \si{\kelvin}: $\sim 8\%$ of \textbf{DPPC} lipids are found in the red phase, which starts nucleating into the blue one ($\sim 92\%$) - see also Supplementary Movie $1$. 
The transition probability between the two phases is also detected and reported on the black arrows.
By using the same setup that detected the gel/liquid separation at $293$ \si{\kelvin}, LENS-based clustering identifies two dominating phases in the \textbf{DPPC} bilayer at $T = 273$ \si{\kelvin} and $T = 323$ \si{\kelvin}, respectively: a cyan domain with lower $\delta_i$ \textit{vs.} a red environment with higher $\delta_i$, respectively (Figure \ref{fig:fig03}b).  Global statistical analysis summarized in Figure \ref{fig:fig03}c by the \textit{Variability} of $D_i^{T}$ distributions reveals that the dynamic reshuffling of lipids is considerably reduced in the cyan domain compared to the red one ($\sim 2-6$ times). This indicates that the lipids into the cyan cluster most probably correspond to the gel phase, while the lipids in the red environment behave as a liquid phase, as also evident in the red disordered lipid tails compared with the more extended/ordered cyan ones (see the snapshot in Figure \ref{fig:fig03}a).
These data thus demonstrate how LENS can blindly distinguish between gel (cyan) and liquid (red) lipid phases and efficiently detect their nucleation and transitions across temperature variations.

For case (ii), we analyze $500$ consecutive frames taken every $\Delta t=0.1$ \si{\nano\second} along  $50$ \si{\nano\second} of MD simulation at $T = 268$ \si{\kelvin} of a periodic box containing $2048$ water molecules in total, $1024$ of which are in the solid state and arranged in a typical hexagonal ice crystal configuration, while the other $1024$, segregated from the first ones, are in the liquid phase (Figure \ref{fig:fig03}d). Shown in Figure \ref{fig:fig03}e, the LENS signals for all water molecules ($\delta_i(t)$ data) clearly demonstrate the presence of two main phases coexisting: one corresponding to low $\delta_i$ values (more static behavior), while the second one characterized by higher $\delta_i$ values (more dynamic). 
HC clustering on the dendrogram reduces the number of clusters (Supplementary Figure $7$a), identifying three main dynamic phases (Figure \ref{fig:fig03}e): the ice phase (in white), the liquid phase (in red), and the water-ice interface (in cyan). 
The interconversion diagram of Figure \ref{fig:fig03}f (right) reveals how the ice and liquid phases exchange molecules through such interface cyan region.
We underline how such a neat classification is typically non-trivial to be attained \textit{via}  sophisticated abstract structural descriptors such as, \textit{e.g.}, SOAP, \cite{capelli2022,offeidanso2022,monserrat2020} and typical pattern recognition algorithms. 
On the other hand, with LENS the detection of different dynamic environments emerges in a straightforward manner and simply by tracking differences in the local reshuffling of the individual water molecules.

\subsection*{Into discrete solid-like dynamics}

As completely different test cases, we also tested LENS on systems with solid-like dynamics. 
In particular, we focused on metal surfaces. While metallic crystals are typically considered hard-matter, it is known that they may possess a non-trivial atomic dynamics even well below the melting temperature.\cite{cioni2022,zepeda2017probing,wang2021atomistic} 
In particular, we consider two Cu FCC surfaces \textbf{Cu(210)} and \textbf{Cu(211)}, having a strikingly different dynamics.

\begin{figure*}[ht!]
\centering
\includegraphics[width=\columnwidth]{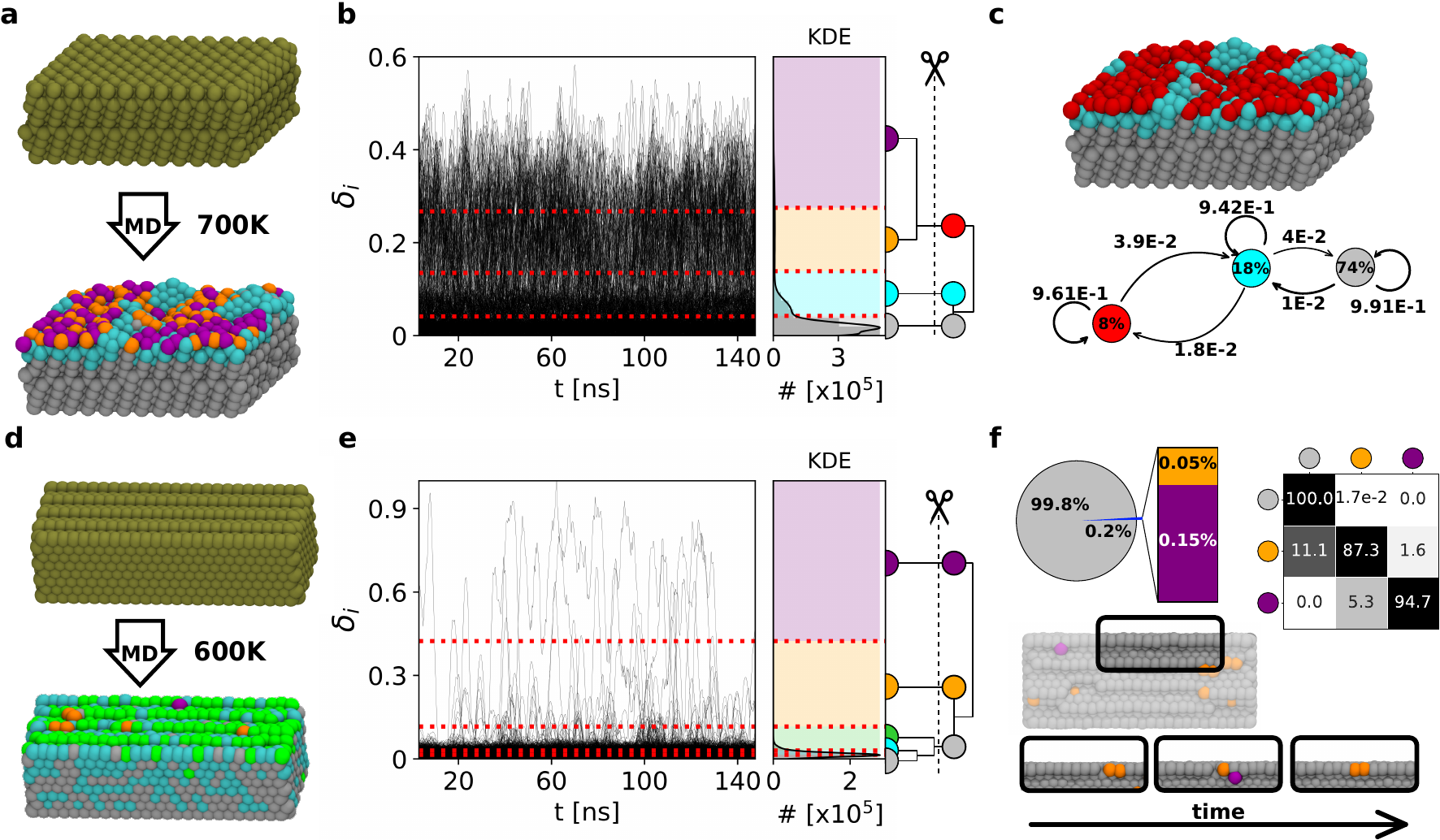}
\caption{LENS Analysis of dynamic metal (Cu) surfaces. (a) MD snapshots of an ideal \textbf{Cu(210)} surface (top: $0$ \si{\kelvin}) and of the same surface at $T = 700$ \si{\kelvin} (bottom): atoms colored according to their LENS-detected dynamic environments of belonging. (b) Time-series of LENS signals, $\delta_i(t)$, with the KDE of LENS distribution, and interconnection dendrogram. Four dynamic domains are first identified by KMeans and then merged into three clusters via HC. (c) MD snapshot of \textbf{Cu(210)} stable bulk in gray, surface in cyan, and dynamic surface spots in red (top). Dynamic interconversion diagram reports the transition probabilities on the arrows and the cluster composition percentages within the colored circles (bottom). (d) MD snapshots of \textbf{Cu(211)} ideal (top) and equilibrated surface at $T = 600$ \si{\kelvin} (bottom) colored according to LENS clusters. (e) Time-series of LENS signals,  $\delta_i(t)$, with the KDE of LENS distribution, and interconnection dendrogram. Five clusters are detected by the LENS-based analysis and merged into three macroclusters. (f) Pie-chart of the clusters compositions and transition probability matrix of the clusters (top). The merged clusters define the surface characterization: the bulk (silver domain), and dynamic atoms which move on the surface breaking/reconstructing rows (orange and purple). Representative MD snapshots showing the surface reconstructions over time are shown on the bottom.}
\label{fig:fig04}
\end{figure*}

We use a $150$ \si{\nano\second} long atomistic MD trajectory of a \textbf{Cu(210)} composed of $2304$ Cu atoms at $T = 700$ \si{\kelvin} (Figure \ref{fig:fig04}a) conducted with a dynamically-accurate deep-potential neural network force field trained on DFT calculations.\cite{cioni2022} 
We analyze with LENS $502$ consecutive frames taken every $\Delta t = 0.3$ \si{\nano\second} along the MD simulation (see Methods section for details).
The LENS signals indicate that the large part of the atoms of this surface is substantially static, while a considerable fraction of the atoms is more dynamic. 
The LENS-based clustering, applied coherently with the protocol described above, detects three main dynamic domains (Figure \ref{fig:fig04}b, right), corresponding essentially to dynamic surface domains (in red), more static surface and sub-surface domains (cyan), and the crystalline bulk of \textbf{Cu(210)} (gray), containing respectively $\sim 8\%$, $\sim 18\%$, and $\sim 74\%$ of the Cu atoms in the model system (Figure \ref{fig:fig04}c: cluster populations in the colored circles). 
The dynamic interconversion plot in Figure \ref{fig:fig04}c reports the probabilities (in $\Delta t = 0.3$ \si{\nano\second}) for atomic exchange between the three main LENS environments, revealing a continuous dynamic exchange of atoms between surface, sub-surface and bulk in the nanosecond-scale consistent with what recently demonstrated.\cite{cioni2022}

As a second case, we analyze a \textbf{Cu(211)} surface composed of $2400$ atoms at $600$ \si{\kelvin} (Figure \ref{fig:fig04}d). We analyze with LENS $502$ consecutive frames taken every $\Delta t = 0.3$ \si{\nano\second} along an MD simulation performed with the same deep-potential force field of the previous case (see Methods for details).\cite{cioni2022} 

Such \textbf{Cu(211)} surface has completely different dynamics than the \textbf{Cu(210)} one. In this case, the time-series $\delta_i(t)$ data provide clear evidence of strikingly non-uniform dynamics (Figure \ref{fig:fig04}e). 
In the \textbf{Cu(210)} simulation at $700$ \si{\kelvin} LENS shows a "fluid-like" atomic surface dynamics. Conversely, in the \textbf{Cu(211)} surface the LENS-based clustering shows that most of this surface is solid/static (Figure \ref{fig:fig04}f: $\sim 99.8\%$ of atoms in the gray cluster and have a low $\delta_i$), while sparse atoms (Figure \ref{fig:fig04}f: $\sim 0.1$-$0.2\%$ in the orange and violet clusters) diffuse and move fast on the surface (large $\delta_i$ LENS signal). 
Such sparse atoms dynamically emerge, diffuse, and reabsorb on the \textbf{Cu(211)} surface in a dynamic fashion: in total, we observe $\sim 200$ gray-to-orange transitions over $\sim 500$ sampled frames (transition frequency of one event every $750$ \si{\pico\second} of simulation).
The transition matrix in Figure \ref{fig:fig04}f describes the kinetic hierarchy between the different static/dynamic LENS states, revealing in orange those atoms in the surface edges which are prone to move (Figure \ref{fig:fig04}f, bottom: MD snapshot), while in violet are the atoms moving at high-speed on the surface after leaving the orange edge defects (see also Supplementary Movie $2$).

In this last case, LENS reveals a strikingly non-uniform dynamics governed by local rare fluctuations, which are typically poorly captured by average-based analyses such as, \textit{e.g.}, pattern recognition approaches, or the global statistical analysis reported for the previous cases (Supplementary Figure $12$a).\cite{cioni2022,rapetti2022} 
This underlines the efficiency of a local time-lapse LENS analysis to detect such rare fluctuations, which has been challenged further with other prototypical case studies as discussed below.

\subsection*{LENS detection \& tracking of local fluctuations}

We tested LENS on other molecular systems whose dynamics is dominated by local fluctuations.

\begin{figure*}[ht!]
\centering
\includegraphics[width=\columnwidth]{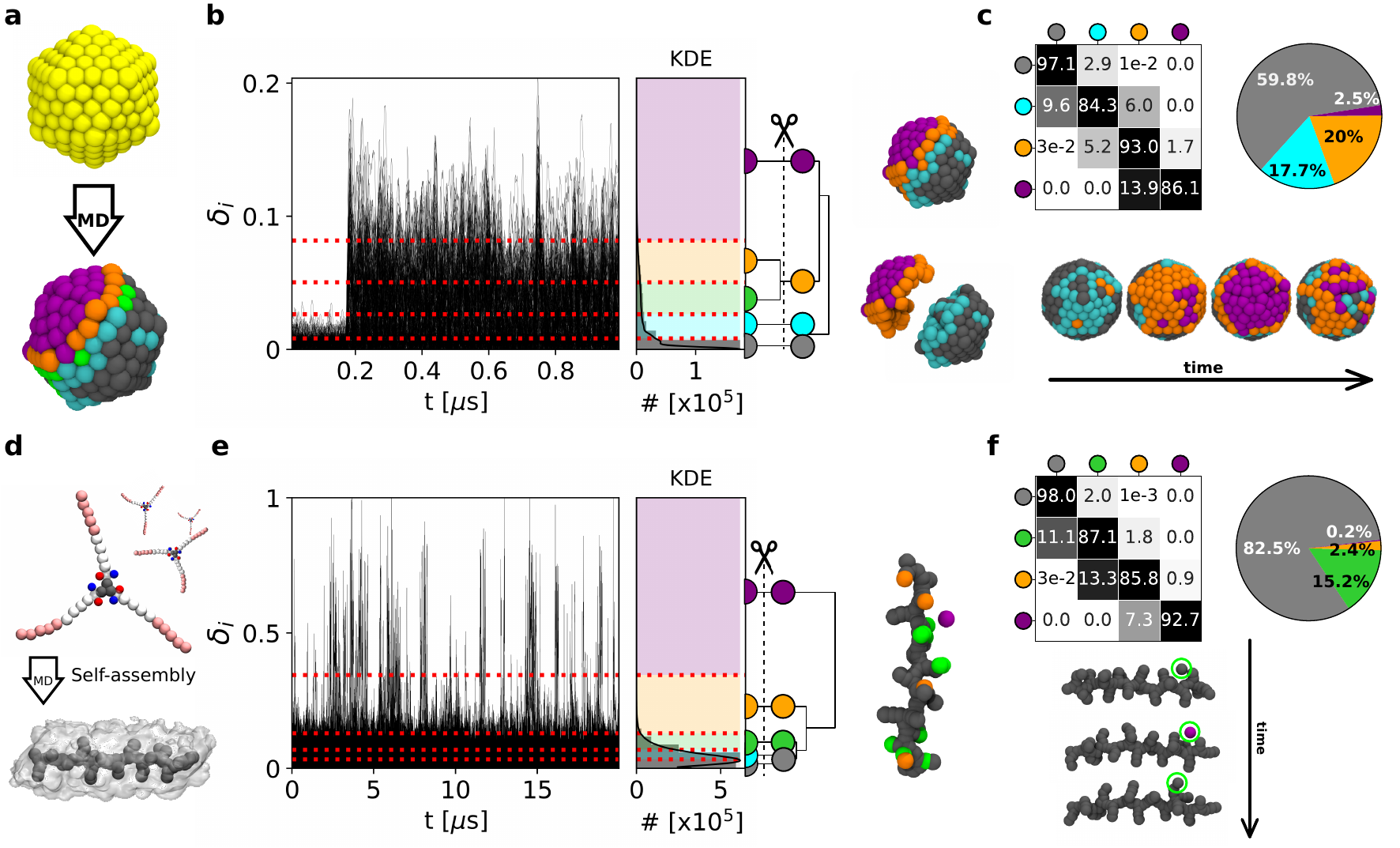}
\caption{LENS analysis for discrete-like dynamics and local fluctuations. (a) Ideal icosahedral \textbf{Au-NP} (top: at $0$ \si{\kelvin}) and at $200$ \si{\kelvin} (bottom): atoms colored based on their LENS clusters of belonging. (b) LENS analysis: time-series $\delta_i(t)$ signals (left), with the KDE of LENS distribution and interconnection dendrogram (center). Right: The four HC resulting LENS clusters show a clear characterization of the \textbf{Au-NP}: one ordered/static region (gray), one intermediate ordered/dynamic domain (cyan), and a mobile area (in orange and purple). (c) Transition probability matrix and cluster composition pie chart (top). Bottom: example of local symmetry breakage in the icosahedral \textbf{Au-NP}. After $\sim180$ \si{\nano\second} of MD simulation, between the 2nd and 3rd snapshots from the left, one vertex (in orange: natively having $5$ neighbor atoms) disappears and is replaced by a rosette (in violet: having $6$ neighbor atoms). (d) \textbf{BTA} monomers (top) and an equilibrated model of a \textbf{BTA} self-assembled fiber (bottom). (d) LENS analysis: time-series $\delta_i(t)$ data (left), with related KDE of LENS distribution and interconnection dendrogram (center). Right: detected LENS clusters, corresponding to the bulk (in gray) and the defect domains in the \textbf{BTA} fiber (green and orange), and to the monomers diffusing from defect to defect on the fiber surface (in purple). (f) Transition probability matrix and cluster population pie-chart (top). Bottom: example of monomer motion (in the green circle) between defects on the fiber surface, consistent with what the processes of monomer reshuffling demonstrated recently for these fibers.\cite{bochicchio2017natcomm,gasparotto2020,gardin2022}}
\label{fig:fig05}
\end{figure*}

First, we focus on a $309$-atoms icosahedral Gold nanoparticle (Figure \ref{fig:fig05}a: \textbf{Au-NP}). It is known that such metal NPs may possess non-trivial dynamics even at room temperature.\cite{rapetti2022} 
We analyze $1000$ consecutive frames taken every $\Delta t = 1$ \si{\nano\second} along $1$ \si{\micro\second} of MD simulation at $200$ \si{\kelvin} of temperature (all atoms are thermalized to guarantee that the temperature is globally constant in the \textbf{Au-NP} -- see Methods for details).\cite{rapetti2022} 
At $T = 200$ \si{\kelvin}, the atomic motion is reduced and the ideal icosahedral architecture of the \textbf{Au-NP} is consequently more stabilized than at, \textit{e.g.}, room temperature.\cite{rapetti2022} 
Nonetheless, after $\sim 180$ \si{\nano\second} of MD simulation the LENS signal rapidly increases from $\sim 0.02$ to $\sim 0.18$ (Figure \ref{fig:fig05}b: $\delta_i(t)$). 
HC clustering of the dendrogram of Figure \ref{fig:fig05}b provides four main LENS dynamic domains (in gray, cyan, orange, and violet, going from the lowest to the highest $\delta_i$ values). Focusing on one \textbf{Au-NP} vertex (Figure \ref{fig:fig05}c, bottom: in the \textbf{Au-NP} center), its surrounding area, initially static (in gray in the 1st MD snapshot on the left), becomes this vertex becomes suddenly more dynamic (2nd MD snapshot: in orange) and, as a dynamic wave, this area turns then violet (3rd snapshot).
Between the 2nd and 3rd snapshots from the left in Figure \ref{fig:fig05}c (bottom), LENS detects a local event well-known in icosahedral Au NPs: one vertex (having five-neighbors in an ideal icosahedron) penetrates inside the NP surface generating a concave "rosette" (having six-neighbors -- in violet).\cite{apra2004}
Such local transition/fluctuation breaks-down the \textbf{Au-NP} symmetry, generating a dynamic region that then coexists with a more static area, in gray (see also Supplementary Movie $3$).
The data in Figure \ref{fig:fig05}c (top) report the transition probabilities between the detected LENS dynamics domains.
This case demonstrates how rare local fluctuations may generate larger collective rearrangements and the efficiency of LENS in detecting them. 

Local transitions/fluctuations are not exclusive of crystalline-like materials, but may be present also in soft systems. We use LENS to analyze a water-soluble 1,3,5-benzenetricarboxamides (\textbf{BTA}) supramolecular polymer composed of monomers that self-assemble directionally \textit{via} $\pi-\pi$ stacking and hydrogen-bonding interactions (Figure \ref{fig:fig05}d).\cite{garzoni2016,leenders2013} 
It has been demonstrated how these supramolecular fibers possess interesting dynamics due to defects that continuously form and annihilate in a dynamic way in the monomer stack.\cite{bochicchio2017natcomm,gardin2022,gasparotto2020}
In this case, we analyze  $20001$ consecutive frames taken every $\Delta t = 1$ \si{\nano\second} along $20$ \si{\micro\second} of CG-MD simulation at room temperature (see Methods for details).\cite{bochicchio2017natcomm,gasparotto2020} 
Recently, unsupervised clustering of SOAP data extracted from the MD trajectories of such \textbf{BTA}-fibers allowed the unbiased detection of the fiber's defects. However, reconstructing \textit{a posteriori} from such structural data the dynamics of these defects and of monomers' diffusion between them is non-trivial.\cite{gasparotto2020,gardin2022} 
Nonetheless, the time-series $\delta_i(t)$ data in Figure \ref{fig:fig05}e clearly show how the dynamics of such fibers is strongly controlled by sharp local fluctuations that are well captured by LENS. HC clustering of the LENS data distinguishes well the interior of the fiber as a more static environment (Figures \ref{fig:fig05}e,f: gray cluster), the defects along the fiber as slightly more dynamic (green and orange), and also the monomers diffusing on the fiber surface (in violet).\cite{bochicchio2017natcomm,gasparotto2020}
The transition matrix and pie-chart of Figure \ref{fig:fig05}f show how the gray, green, and orange clusters include the majority of the \textbf{BTA}-monomers. 
On the other hand, sparse monomers ($\sim 0.2\%$) belonging to the violet cluster undergo sharp transitions and instantaneous reshuffling of their local neighbors.
These are the monomers that are diffusing defect-to-defect on the fiber surface, which provides a picture of the internal dynamics of such complex \textbf{BTA} fibers in optimal agreement with previous studies.\cite{bochicchio2017natcomm,gasparotto2020,gardin2022,bochicchio2018mol} 

Also in these cases (as in the \textbf{Cu(211)} surface of Figure \ref{fig:fig04}d-f), LENS is found efficient in detecting and tracking local fluctuations that play a dominant role in the dynamics of the entire system.
It is worth noting how in all such cases a time-independent (pattern recognition-based) statistical analysis of neighbors variability is inefficient to reconstruct such non-uniform dynamics, due to the low statistical weight of the local events occurring in these systems (see also Supplementary Figure $12$).

\section*{Discussion}

Many molecular systems are controlled by local fluctuations that are often difficult to detect and typically lost in average-based analyses. Here we present a new general descriptor designed to track local fluctuations in complex dynamic systems, named Local Environments and Neighbors Shuffling (LENS). Different from many descriptors, LENS is based on the concept of neighbor identities (IDs) instead of, \textit{e.g.}, molecular/atomic species. At each sampled time-frame along a trajectory, our analysis builds a string listing the neighbor IDs surrounding each particle $i$ in the system. Within the time-interval between consecutive time-frames, LENS measures the variations in the neighbor IDs in terms of addition, subtraction, or reshuffling of neighbors (Figure \ref{fig:fig01}). Large time-lapse variations in the local neighborhood provide strong LENS signals, while weak LENS signals indicate reduced dynamics in the local environment surrounding a given particle $i$.

We tested LENS in a number of systems with strikingly different internal dynamics. Shown in Figure \ref{fig:fig02}, LENS reveals that a bicomponent lipid bilayer is characterized by surface patches, with different molecular reshuffling dynamics, which correspond to the segregation of the lipid species into two domains. In Figure \ref{fig:fig03}, we demonstrate how our time-series LENS analysis detects efficiently phase transitions and coexistence of different phases: \textit{e.g.}, in a \textbf{DPPC} lipid bilayer undergoing gel-to-liquid transition increasing the temperature from $273$ \si{\kelvin} to $323$ \si{\kelvin}, or in a liquid water-ice system at freezing/melting temperature. 

When a system is characterized by statistically-dominant dynamic domains, the time-dependent LENS and global (time-independent) statistical analyses correlate (Figure \ref{fig:fig02} and Figures \ref{fig:fig03}a-c). Conversely, system dynamics dominated by rare local fluctuations are poorly  described by global statistical analyses (Supplementary Figure $12$).
In the \textbf{Cu(211)} surface (Figure \ref{fig:fig04}d-f), for example, a global statistical analysis based on a pattern recognition approach identifies only one domain, as reported in Supplementary Figure $12$a, meaning that the sparse atoms diffusing fast on the metal surface are not statistically relevant and are statistically-lost in such analyses.
Rare local transitions are not captured by a global time-independent analysis even in the systems of Figure \ref{fig:fig05}. 
This is not necessarily an exclusive problem of time-independent analyses conducted with this specific descriptor: also other descriptors such as, \textit{e.g.}, SOAP, coordination number, etc., are in fact efficient as far as they are used to detect statistically-relevant dynamic/structural populations and patterns.
Nonetheless, the results of Figures \ref{fig:fig04}d-f and \ref{fig:fig05} demonstrate how a local time-dependent LENS analysis is efficient in detecting and tracking such local fluctuations, and in this sense appears as more general, complete, and robust than an average time-independent investigation. 
In addition, while average-based and global pattern recognition analyses work typically well when one knows what to search, this is less the case for LENS.
The LENS analysis in fact only requires knowing the IDs of the interacting particles and having a sufficiently sampled trajectory.
This is fundamental in most practical cases where the nature of a system is not known \textit{a priori}.

LENS has also some intrinsic limitations. Based on its definition, if in $\Delta t$ the neighbors do not change (same IDs) but move remaining in the $r_{cut}$ sphere (local structural rearrangement of the neighborhood), LENS provides no signal. This is opposed to descriptors such as, \textit{e.g.}, SOAP that -- being permutationally invariant -- provide \textit{vice versa} a signal in case of local rearrangements, but no signal in case of a permutation of IDs (keeping the same structural displacement). This makes LENS best suited to measure local dynamicity rather than local structural variations, which is nonetheless key in many complex systems where dynamics plays a major role.
At the same time, one key advantage of LENS is its abstract definition. This makes it well suited to analyze a variety of trajectories of systems for which the identities of the moving units are known and, in principle, not necessarily restricted to molecular ones. 

\section*{Methods}
\subsection*{MD simulations}

All data concerning the molecular models and the MD trajectories analyzed herein are available at: \url{https://github.com/GMPavanLab/LENS} (this link will be replaced with a definitive Zenodo archive upon acceptance of the final version of this paper). 

The \textbf{DIPC}/\textbf{DPPC} lipid bilayer (Figure \ref{fig:fig02}) is simulated using the Martini2.2 force field.\cite{marrink2007} A binary mixture of dipalmitoyl-phosphat-idylcholine (\textbf{DPPC}) and dilinoleoyl-phosphatidyl-choline (\textbf{DIPC}), with $2$:$3$ molar ratio, is used to model the coexistence of liquid-crystalline and gel phases into such self-assembled bilayer.
To get the separation of the bilayer into domains of coexisting phases, the mixture was simulated at $T = 280$ \si{\kelvin}.
The initial configuration of the binary lipid mixture in water is generated using \emph{insane} \cite{wassenaar2015computational} with the specified box dimensions ($18$ x $18$ x $11$ \si{\nano\meter}).
The bilayer system is composed of $1150$ lipids, consisting of $2$:$3$ \textbf{DIPC}:\textbf{DPPC} on each leaflet, and $17987$ (W) water molecules.
To prevent water crystallization ($T < 290$ \si{\kelvin} in Martini),\cite{marrink2007} $\sim 5\%$ of regular water particles are substituted by the anti-freezing water particles.
For non-bonded interactions, a reaction-field electrostatics algorithm is used with a Coulomb cutoff of $r_c=1.1$ \si{\nano\meter} and a dielectric constant of $15$.
The cutoff for Lennard-Jones interactions is set to $r_{LJ}=1.1$ \si{\nano\meter}.
The timestep used during the MD simulation is $\delta t = 20$ \si{\femto\second}.
The system is preliminarily minimized and equilibrated for $t = 100$ \si{\nano\second}.
A production run is then performed for $t=15$ \si{\micro\second}, and the data acquisition is performed every $1$ \si{\nano\second}.
The solvent and membrane are coupled separately using a v-rescale thermostat with a relaxation time of $t = 1.0$ \si{\pico\second}.
During the equilibration, the pressure is maintained at $p=1$ \si{\bar} using the Berendsen barostat with the semi-isotropic coupling scheme, a time constant of $\tau_p=4$ \si{\pico\second}, and compressibility $c = 3*10^{-4}$ \si{\per\bar}. During the production, the Parrinello-Rahman barostat is used, with a time constant of $\tau_p=12$ \si{\pico\second}.
An equilibrium part of the trajectory is analyzed (the last $10$ \si{\micro\second}) every $\Delta t = 10$ \si{\nano\second} ($1001$ sampled frames).

The bicomponent \textbf{F-NP}/\textbf{H} micelle (Supplementary Figure $2$) was simulated at $T = 300$ \si{\kelvin} in explicit water \emph{via} Martini2.2 \cite{marrink2007} scheme (see reference \cite{cardellini2022} for further details). 
The system is a binary mixture of p-nitrophenyl ester of n-stearoyl L-phenylalanine (\textbf{F-NP}) and  n-stearoyl L-histidine (\textbf{H}) with $1$:$1$ molar ratio (N$_{\textbf{F-NP}}=100$ and N$_{\textbf{H}}=100$).
The initial configuration consists of N$_{\textbf{F-NP}}=100$ and N$_{\textbf{H}}=100$ randomly dispersed surfactants, which assemble into a single micelle within a $10$ \si{\micro\second} long MD simulation sampled every $1$ \si{\nano\second}.
The last $3$ \si{\micro\second} of the MD trajectory is considered representative of the equilibrium\cite{cardellini2022} and used for the analysis -- $3001$ analyzed frames taken every $\Delta t = 1$ \si{\nano\second} along the MD.

All the \textbf{DPPC} lipid bilayer trajectories at $T = 293$ \si{\kelvin}, $273$ \si{\kelvin} and $323$ \si{\kelvin} (Figure \ref{fig:fig03}a-c) are obtained from MD simulations of a bilayer model composed of N$_{DPPC}=1152$ \textbf{DPPC} lipids, simulated and parameterized in explicit water \emph{via} Martini2.2, \cite{marrink2007} as reported in reference \cite{capelli2021}.
The equilibrated-phase MD trajectories used for the analyses are in all cases 1 \si{\micro\second}. A total of $1001$ frames extracted every $\Delta t = 1$ \si{\nano\second} along the MD trajectories are used for the analyses.

The atomistic Ice/Water interface model of Figure \ref{fig:fig03}d-f is simulated employing the direct coexistence technique. The \textbf{TIP4P}/\textbf{Ice} water model\cite{abscal2005} is used to model both the solid phase of ice $I_h$ and the phase of liquid water. The direct coexistence technique is based on the idea to put in contact more phases in the same box and at constant pressure. To get the coexistence, the temperature is set at $T = 268$ \si{\kelvin} (the energy is constant at $268$ \si{\kelvin} and the system melts at $269$ \si{\kelvin}\cite{garcia2006melting}), kept constant using the v-rescale thermostat with a relaxation time of $t=0.2$ \si{\pico\second}.
The initial configuration of the ice $I_h$ is obtained using the \emph{Genice} tool proposed by Matsumoto \emph{et al.}\cite{matsumoto2018} generating a hydrogen-disordered lattice with zero net polarization satisfying the Bernal-Fowler rules.
To equilibrate the solid lattice, anisotropic \emph{NPT} simulation is carried out using the c-rescale barostat, with a time constant of $\delta t =20$ \si{\pico\second} and compressibility of $9.1*10^{-6}$ \si{\per\bar}. The equilibration lasted $10$ \si{\nano\second} at ambient pressure (1 atm).
The liquid phase is obtained from the same ice $I_h$ solid phase, performing a \emph{NVT} simulation at $T = 400$ \si{\kelvin} to quickly melt the ice slab.
Thus, both the solid and liquid phases are obtained with the same number of molecules ($1024$) and box dimensions.
The liquid phase is then equilibrated at $T = 268$ \si{\kelvin} for $t=10$ \si{\nano\second}, using the c-rescale barostat in semi-isotropic conditions and compressibility of $c=4.5*10^{-5}$ \si{\bar}.
The two phases are, then, put in contact and equilibrated for $t=10$ \si{\nano\second} using the c-rescale pressure coupling with the water compressibility  ($c=4.5*10^{-5}$ \si{\bar}) at ambient pressure.
The production \emph{NPT} ice/water coexistence MD simulation (Figure \ref{fig:fig03}d-f) is performed in semi-isotropic conditions, with the pressure applied only in the direction perpendicular to the ice/water interface. This allows to reproduce the strictly correct ensemble for the liquid-solid equilibrium simulation by the direct coexistence technique.
After the equilibration, a production run is performed for $t=50$ \si{\nano\second}, sampled and analyzed every $0.1$ \si{\nano\second}. 
All the trajectories analyzed for the systems simulated above are obtained using the GROMACS software.\cite{Hess2008}

The atomistic models of the \textbf{Cu(210)} and \textbf{Cu(211)} surfaces (Figure \ref{fig:fig04}) are composed of N$_{210}=2304$ and N$_{211}=2400$ atoms, respectively. The MD simulations are conducted at $T=700$ \si{\kelvin} and at $T=600$ \si{\kelvin} respectively for the two example surfaces. Deep-potential MD simulations of both Cu surfaces are conducted with the LAMMPS software\cite{thompson2022} using a Neural Network potential built using the DeepMD platform,\cite{wang2018} as described in detail in reference \cite{cioni2022}.
The sampled trajectories are $150$ \si{\nano\second} long. A total of $502$ frames are extracted every $\Delta t=0.3$ \si{\nano\second} along the MD trajectories and used for the LENS analyses.

The atomistic model for the icosahedral \textbf{Au-NP} is composed of N$_{Au-NP}=309$ gold atoms (Figure \ref{fig:fig05}a-c). The \textbf{Au-NP} model is parametrized according to the Gupta potential, \cite{gupta1981} and is simulated for $1$ \si{\micro\second} of MD at $T=200$ \si{\kelvin} using the LAMMPS software\cite{thompson2022} as described in detail in reference \cite{rapetti2022}.
$1000$ frames are extracted every $\Delta t = 1$ \si{\nano\second} of the MD trajectory and then used for the analyses.

The coarse-grained \textbf{BTA} fiber model is built consistent with the MARTINI force field\cite{marrink2007} and optimized as described in detail in references \cite{bochicchio2017acs,bochicchio2017natcomm}.
In particular, the fiber model is composed of N$_{BTA}=80$ \textbf{BTA} monomers. 
A trajectory of $20$ \si{\micro\second}, obtained with the GROMACS software\cite{Hess2008}, is then analyzed every $\Delta t = 1$ \si{\nano\second} ($20001$ sampled frames in total).

\subsection*{Pre-processing of the trajectories}
All MD trajectories are firstly pre-processed in order to obtain plain \texttt{xyz} files keeping only the coordinates of the particles of interest, \emph{i.e.}, considered during the neighborhood's evaluation, as reported in Supplementary Table $1$.
For example, in the lipid bilayer analyses of Figures \ref{fig:fig02} and \ref{fig:fig03}a-c we considered only the tan PO4 (MARTINI) beads as representative of the "center" position of each lipid molecule in the systems. 
For the micelles of Supplementary Figure $2$, we used the center of mass of the surfactant heads as the centers for the analysis, for the water/ice system (Figure \ref{fig:fig03}d-f), we considered only the Oxygens of the water molecules, for the metal surfaces and \textbf{Au-NP} (Figure \ref{fig:fig04} and Figure \ref{fig:fig05}a-c) information of each atom was retained, while for the \textbf{BTA} fiber, we considered only the center of each monomer core as a reference for the LENS analyses.
In all cases, the analysis is then conducted by building at each sampled timestep strings collecting the neighbor IDs of each unit \textit{i} within a sphere of radius $r_{cut}$ (which is set depending on the system and based on the shape and the minima of the radial distribution functions, g(r)$_m$ -- see Supplementary Table $1$ and Supplementary Figure $1$).

\subsection*{Time-lapse LENS analysis}
The instantaneous $\delta_i$ parameter for each unit $i$ in each model system is calculated over time along the system's trajectory from the $C_i$ strings containing the IDs of the neighbor units calculated at times $t$ and $t+\Delta t$ as reported in Equation \ref{eq:01}.
The analysis is then repeated for all units $i$ at all time-intervals $\Delta t$ sampled along the analyzed trajectories, obtaining the $\delta_i$(t) plots of Figures \ref{fig:fig01}b,\ref{fig:fig02}b,\ref{fig:fig03}a,\ref{fig:fig03}e,\ref{fig:fig04}b,\ref{fig:fig04}e,\ref{fig:fig05}b and \ref{fig:fig05}e.
The $\delta_i$ parameter is normalized such that it gives 0 when the local neighborhood does not change and $1$ when it changes completely at each $\Delta t$. 
To remove the noise from each $\delta_i$(t) signal, we process them by using a Savitzky–Golay filter\cite{savitzky1964}, as implemented in the SciPy python package \cite{virtanen2020}, obtaining smoothed $\langle\delta_i(t)\rangle$ signals.
Each $\delta_i$(t) signal is smoothed using a common polynomial order parameter of $p=2$ (see the example study reported in Supplementary Figure $5$) on a time-window of $100$ frames for the bicomponent \textbf{DIPC/DPPC} lipid bilayer system, the \textbf{F-NP/H} micelle, \textbf{DPPC} lipid and for the water/ice interface. A time-window of $20$ frames is instead used for the quasi-crystalline \textbf{Cu} surfaces, for the gold \textbf{Au-NP} and for the  \textbf{BTA} systems.
In order to simplify the notation, we refer to the $\langle\delta_i(t)\rangle$ signal as $\delta_i$.
After the noise reduction, the clustering of the $\delta_i$ data is performed by means of KMeans algorithm \cite{lloyd1982} implemented in SciPy python package \cite{virtanen2020}.
The KMeans algorithm requires the definition of the number of clusters as an input. To guarantee to start from an excess of micro-clusters, we use the following criterion: when multiple peaks or clear discontinuities are detectable in the distributions of the $\delta_i$ data (distributions on the right of Figures \ref{fig:fig01}b,\ref{fig:fig02}b,\ref{fig:fig03}a,\ref{fig:fig03}e,\ref{fig:fig04}b,\ref{fig:fig04}e,\ref{fig:fig05}b and \ref{fig:fig05}e), we initially set the number of microclusters as twice the number of the peaks/discontinuities. When on the other hand just one peak is visible in the $\delta_i$ distribution, the initial number of microclusters is anyways set to five. 
This guarantees that KMeans detects a wide variety of dynamics. Such microclusters (which difference is in some cases statistically irrelevant) are merged hierarchically only \textit{a posteriori} \emph{via} a single link algorithm based on the transition probabilities reported in the transition matrices and on the related HC interconnection dendrograms (\textit{i.e.}, based on the dynamic correlation between the microclusters).
Cutting the HC dendrogram at different levels equals analyzing the dynamic diversity in the analyzed systems with a different resolution, see Supplementary Figures $6$,$7$,$8$,$9$,$10$ and $11$.

We note that the results shown herein are obtained \textit{via} such a simple iterative supervised clustering approach, which in the cases we discuss in this work was found simple, effective, and little sensitive to the tuning of clustering parameters (thus satisfactory from the robustness and reproducibility point of view). Nonetheless, we underline that other (\textit{e.g.}, unsupervised) clustering approaches could be used for the purpose, although they do not always provide consistent results with each other, and where the tuning of the setup parameters may be non-trivial.  

\subsection*{Global statistical analysis}
Average information on the statistically dominant dynamic domains present in the systems can also be obtained from the global dataset of the $C_i$ as described in the text.
For each $i$ unit, the numbers of the contacts with the other neighbor IDs along the trajectory ($D_i^T$, considering all T sampled frames) are collected from the global $C_i$ dataset (see, \textit{e.g.}, Figure \ref{fig:fig01}c). The contacts data are then organized into a contact matrix where the individual entry $i,j$ indicates the total number of neighboring events between the bead $i$ and $j$ in all sampled time-intervals along the analyzed trajectory (Figures \ref{fig:fig02}e and Supplementary Figure $2$e).
The data of each unit $i$ (each row of the matrix) are then centered on the mean and normalized on the standard deviation.
The matrix is then analyzed \textit{via} Hierarchical Clustering (HC). 
In particular, the normalized contact data are gathered by means of Ward method \cite{ward1963} with Euclidean metric (both implemented in SciPy python package\cite{virtanen2020}), and the number of clusters is determined based on the dominant patterns from the sorted matrix (see, \textit{e.g.}, the matrices of Figure \ref{fig:fig02}e and Supplementary Figure $2$e, right).
The \emph{Variability} ($V$) in the contacts is then defined as the inverse standard deviation of each $i$ unit contact distribution: the more a $i$ unit changes its neighborhood along the trajectory, the lower (and more variable) is its $D_i^T$ count (and \textit{vice versa}). 

\subsection*{Data availability}
Details on the molecular models and on the MD simulations, and additional MD data are provided in the Supplementary Information.
Complete details of all molecular models used for the simulations, and of the simulation parameters (input files, etc.), as well as the complete LENS analysis code, are available at: \url{https://github.com/GMPavanLab/LENS} (this temporary folder will be replaced with a definitive Zenodo archive upon acceptance of the final version of this paper). 

\section*{Acknowledgements}
G.M.P. acknowledges the support received by the European Research Council (ERC) under the European Union’s Horizon 2020 research and innovation program (Grant Agreement no. 818776 - DYNAPOL) and by the Swiss National Science Foundation (SNSF Grant IZLIZ2$\_183336$).

\section*{Author contributions}
G.M.P. conceived this research and supervised the work. M.C. developed the LENS descriptor and performed the analyses. M.C., A.C., and C.C. performed the simulations. All authors analyzed and discussed the results. M.C., A.C., and G.M.P. wrote the manuscript.

\section*{Competing interests statement} 
The authors declare no competing interests.

\bibliography{bibliography}

\end{document}


\maketitle
\newpage

\begin{table}[t]
     \centering
    \begin{tabular}{|c|c|c|c|c|c|c|}
    \hline
        \textbf{SYSTEM} & \textbf{$r_{cut}$ [\si{\angstrom}]} &\shortstack{ \textbf{\# of g(r)$_m$} \\ \textbf{peaks}} & \shortstack{\textbf{Length of}\\ \textbf{MD[ns]}} & \shortstack{\textbf{\# of sampled}\\\textbf{frames} }& \shortstack{\textbf{Sampling}\\\textbf{$\Delta t$ [\si{\nano\second}]}} & \shortstack{\textbf{LENS} \\\textbf{center}} \\ \hline
        \shortstack{DPPC/DIPC \\Lipids} & 16 & 3 & 10000 & 1001 & 10 & PO4 \\\hline
        \shortstack{F-NP/H\\ Micelle} & 16 & 3 & 3000 & 3001& 1 & \shortstack{HEAD \\(CoM)} \\ \hline
        \shortstack{DPPC\\ Lipids 293K} & 16 & 3 & 1000 & 1001 & 1 & PO4 \\\hline
        \shortstack{DPPC\\ Lipids 273K} & 16 & 3 & 1000 & 1001 & 1 & PO4 \\\hline
        \shortstack{DPPC\\ Lipids 303K} & 16 & 3 & 1000 & 1001 & 1 & PO4 \\\hline
        \shortstack{TIP4P/Ice\\ Water} & 7.4 & 3 & 50 & 500 & 0.1 & OW \\ \hline
        \shortstack{Cu(210)\\ 700K} & 4.9 & 3 & 150 & 502 & 0.3 & Cu \\ \hline
        \shortstack{Cu(211)\\ 600K} & 4.9 & 3 & 150 & 502 & 0.3 & Cu \\ \hline
        \shortstack{Au-NP \\200K} & 4.4 & 2 & 1000 & 1000 & 1 & Au \\ \hline
        BTA & 15.8 & 3 & 20000 & 20001 & 1 & \shortstack{BENZ\\ (CoM)} \\ \hline
    \end{tabular}
    \caption{Setup details of all the LENS analyses conducted in this work.}
    \label{tab:tab01}
\end{table}

\begin{figure}
 \includegraphics[width=\columnwidth]{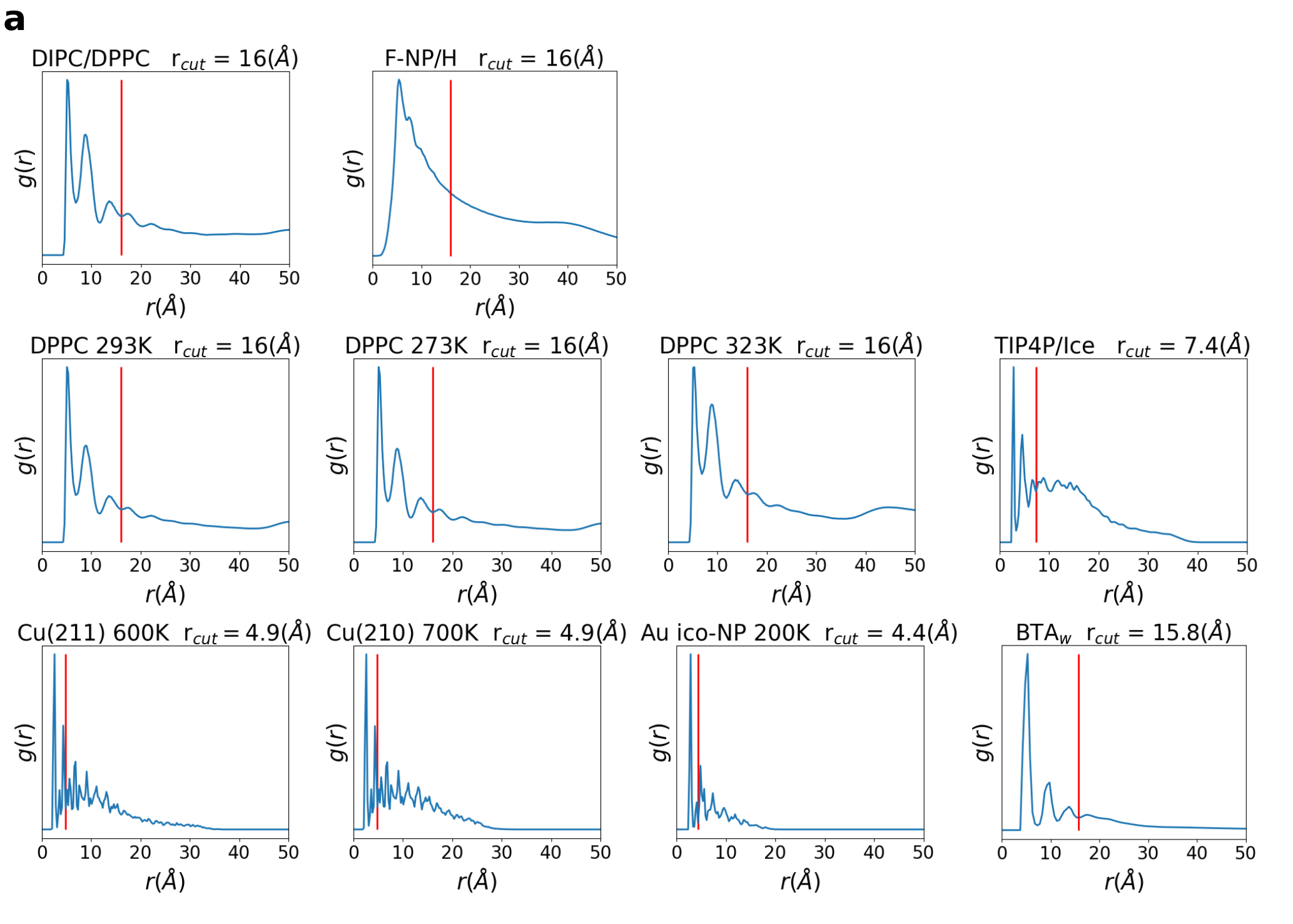}
   \centering
   \caption{(a) Radial distribution functions (g(r)) and cut-off radius $r_{cut}$ for all systems.}
   \label{SIfig01}
\end{figure}

\begin{figure}
 \includegraphics[width=\columnwidth]{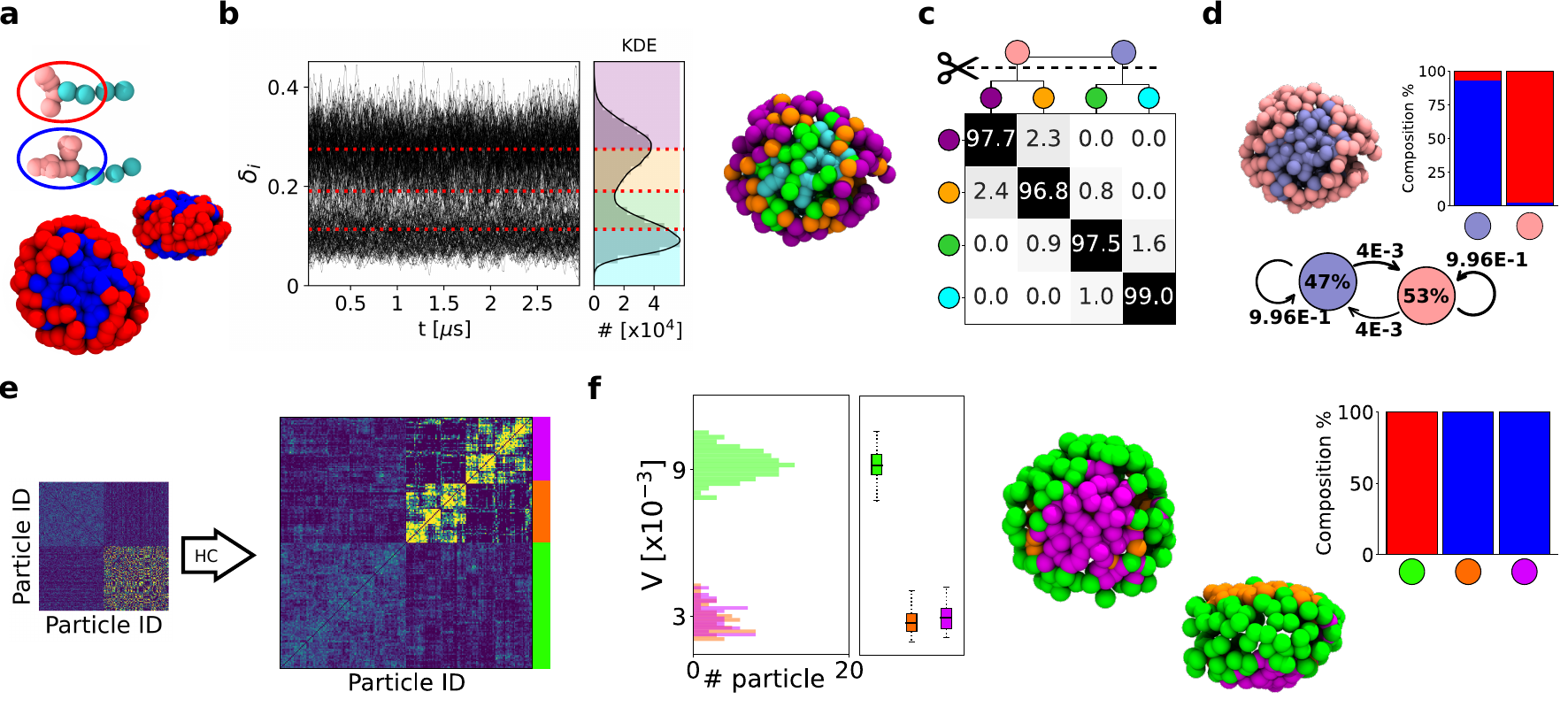}
   \centering
   \caption{LENS analysis of fluid-like systems. (a) Bicomponent amphiphile micelle composed of $100$ \textbf{H} surfactants and $100$ \textbf{F-NP} surfactants colored in red and blue, respectively. (b) Time-series of LENS signals, $\delta_i(t)$, with the Kernel Density Estimate (KDE) of LENS distribution classified into four clusters (left). MD snapshot of the micelle colored according to their clusters of belonging (right). (c) Inter-clusters normalized transition probability matrix. The $p_{ii}$ and $p_{ij}$ matrix entries indicate the $\%$ probability that molecules with LENS signal typical of a cluster $i$ remain in that dynamical environment or move to another one $j$ (with different dynamics) in $\Delta t$. Hierarchical grouping of the  dynamically-closer clusters (dendrogram cutting) is reported on top of the matrix, and it provides two macroclusters, merging cyan and green on one hand, and orange and purple on the other hand. (d) MD snapshot of the micelle colored according to macroclusters in (c): light-blue identifying \textbf{F-NP} surfactants, pink identifying \textbf{H} surfactants (top-left). Cluster composition histogram (top-right) and interconversion diagram (bottom) with the transition exchange probabilities and the cluster population percentages (within colored circle). (e) HC analysis of the $D^{T}$ matrix identifying three main clusters (green, purple, orange). (f) \emph{Variability}, $V$, analysis of the clusters: distributions, median (first and third quartiles), maximum and minimum values (whiskers). The green have higher $V$ than the orange and magenta clusters (left). MD snapshot front and lateral view of the micelle colored according the HC clustering of $D^{T}$ matrix (middle). Cluster composition histogram (top-right): the green cluster is made of \textbf{H} surfactants (in red in (a)), while the orange and magenta ones correspond to the \textbf{F-NP} surfactants (in blue in (a)).}
   \label{SIfig02}
\end{figure}

\begin{figure}
 \includegraphics[width=\columnwidth]{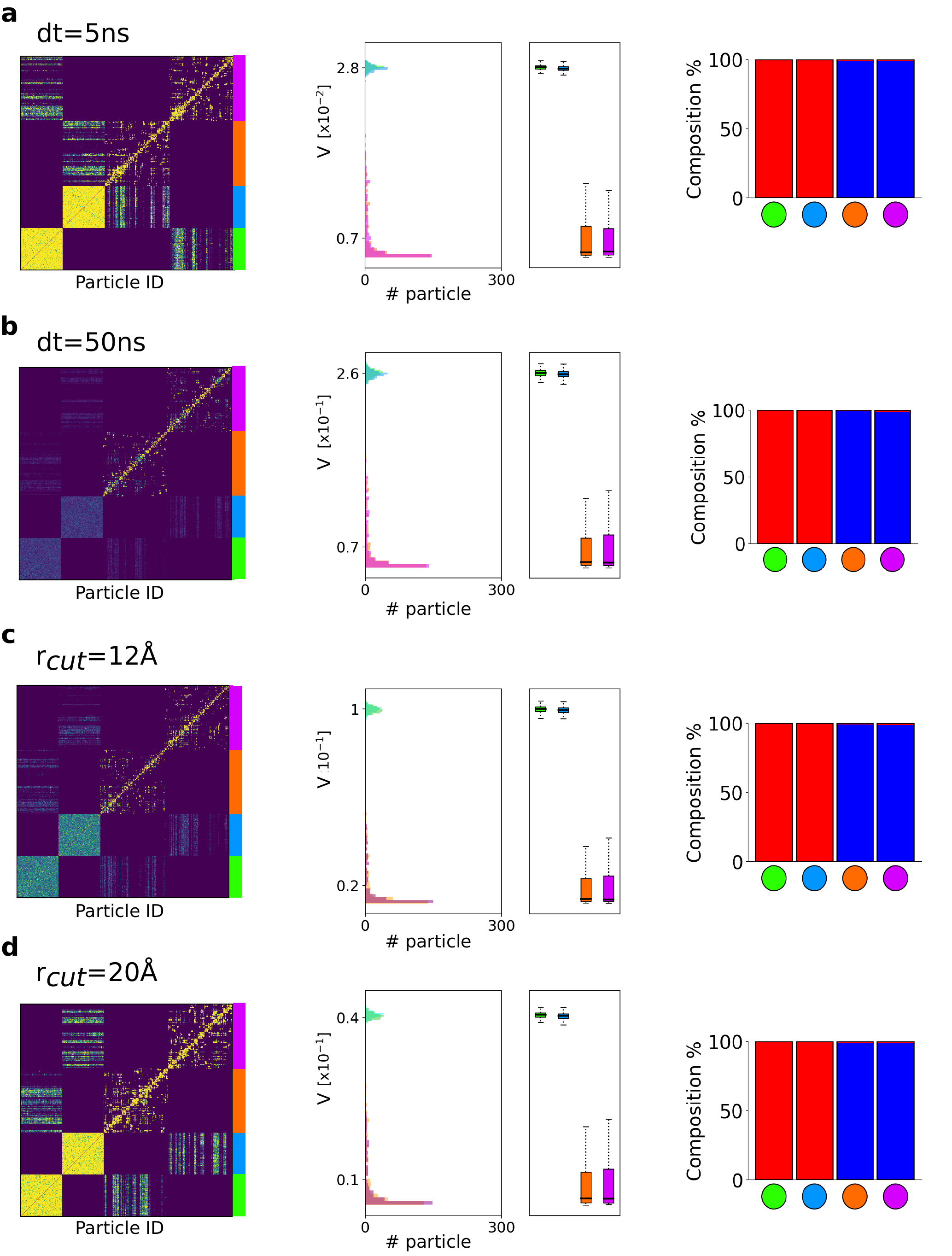}
   \centering
   \caption{Statistical analysis for \textbf{DIPC}/\textbf{DPPC} lipid bilayer, varying sampling step $\Delta t$ or neighborhood cutoff radius r$_{cut}$ while keeping all the other parameters as reported in Table \ref{tab:tab01}. (a) $\Delta t = 5$ \si{\nano\second}, (b) $\Delta t = 50$ \si{\nano\second}, (c) $r_{cut}=12$ \si{\angstrom} and (d) $r_{cut}=20$ \si{\angstrom}.}
   \label{SIfig03}
\end{figure}

\begin{figure}
 \includegraphics[width=\columnwidth]{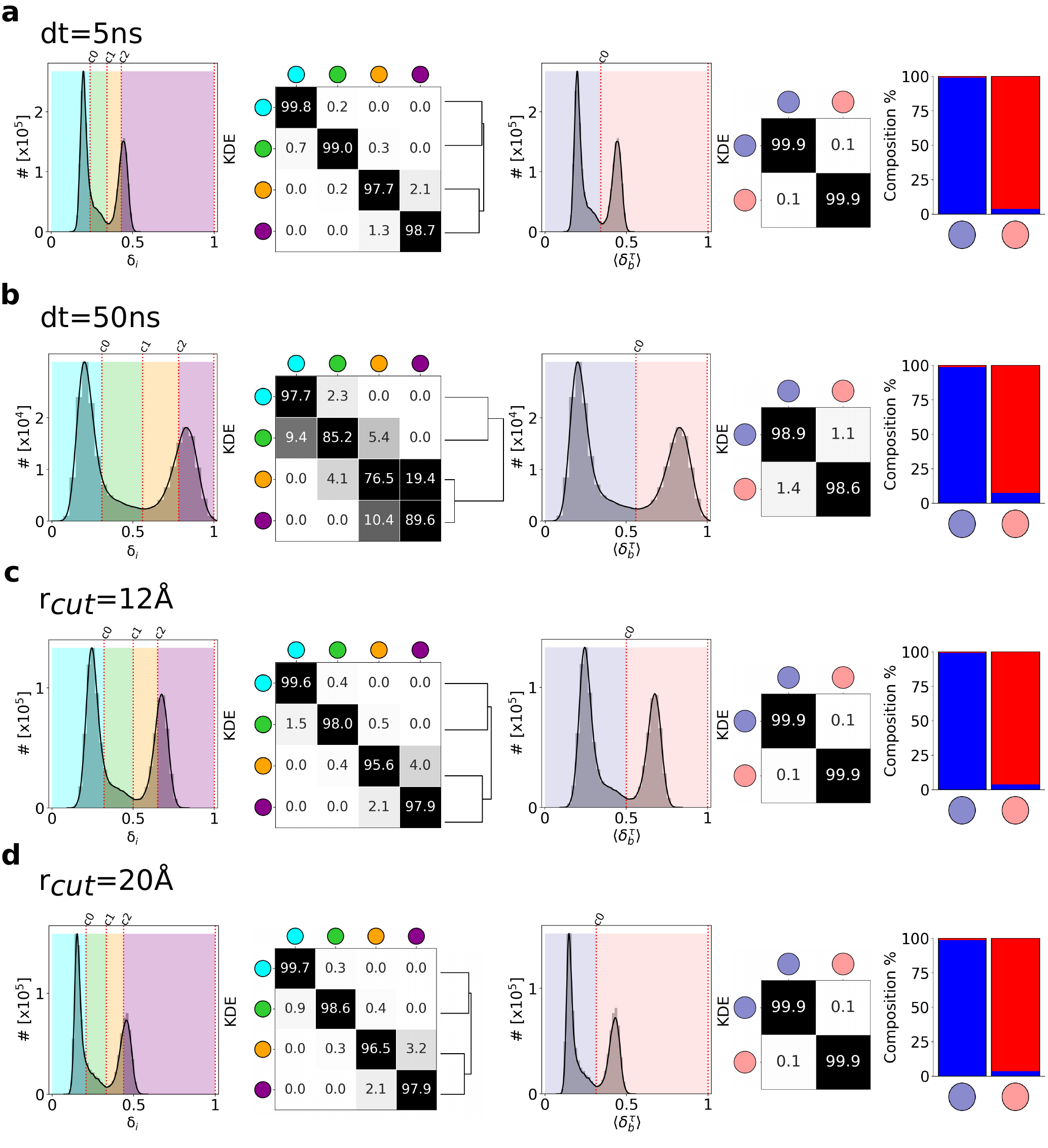}
   \centering
   \caption{LENS analysis for \textbf{DIPC}/\textbf{DPPC} lipid bilayer, varying sampling step $\Delta t$ or neighborhood cutoff radius $r_{cut}$ while keeping all the other parameters as reported in Table \ref{tab:tab01}. (a) $\Delta t = 5$ \si{\nano\second}, (b) $\Delta t = 50$ \si{\nano\second}, (c) $r_{cut}=12$ \si{\angstrom} and (d) r$_{cut}=20$ \si{\angstrom}.}
   \label{SIfig04}
\end{figure}

\begin{figure}
 \includegraphics[width=0.95\columnwidth]{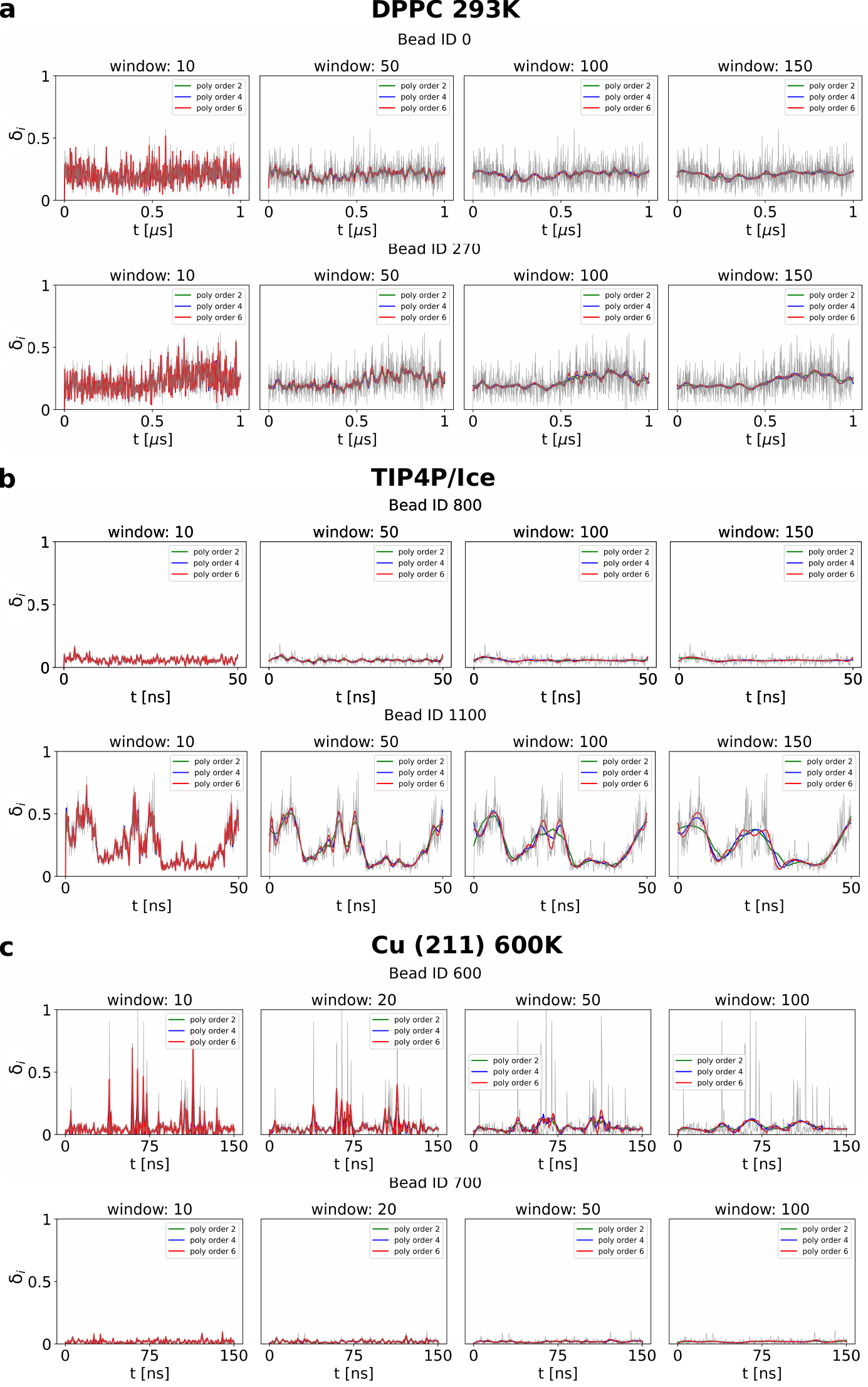}
   \centering
   \caption{Parameter study for applying the Savitzky–Golay filter varying both window interval and polynomial order for two bead examples each of (a) DPPC $293$ \si{\kelvin}, (b) \textbf{TIP4P/Ice} water and (c) \textbf{Cu (211)} copper slab.}
   \label{SIfig05}
\end{figure}

\begin{figure}
 \includegraphics[width=\columnwidth]{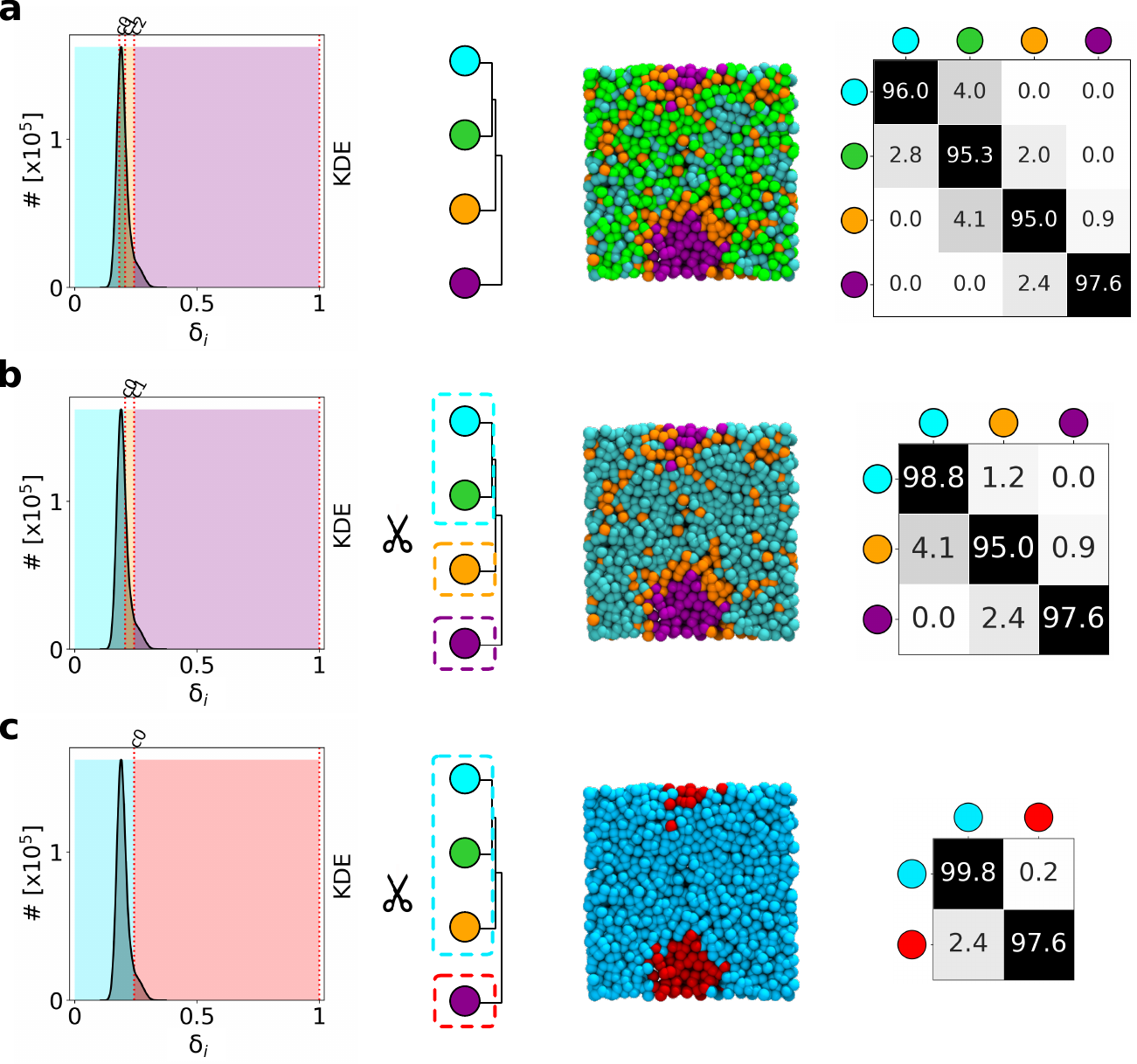}
   \centering
   \caption{\textbf{DPPC} Lipid bi-layer at $T=293$ \si{\kelvin}: merging LENS clusters at different levels (no merging (a), two clusters merged into one (b) and three clusters merged into one (c)) following the hierarchy given by the dendrogram, example snapshot and transition probability matrix.}
   \label{SIfig06}
\end{figure}

\begin{figure}
 \includegraphics[width=\columnwidth]{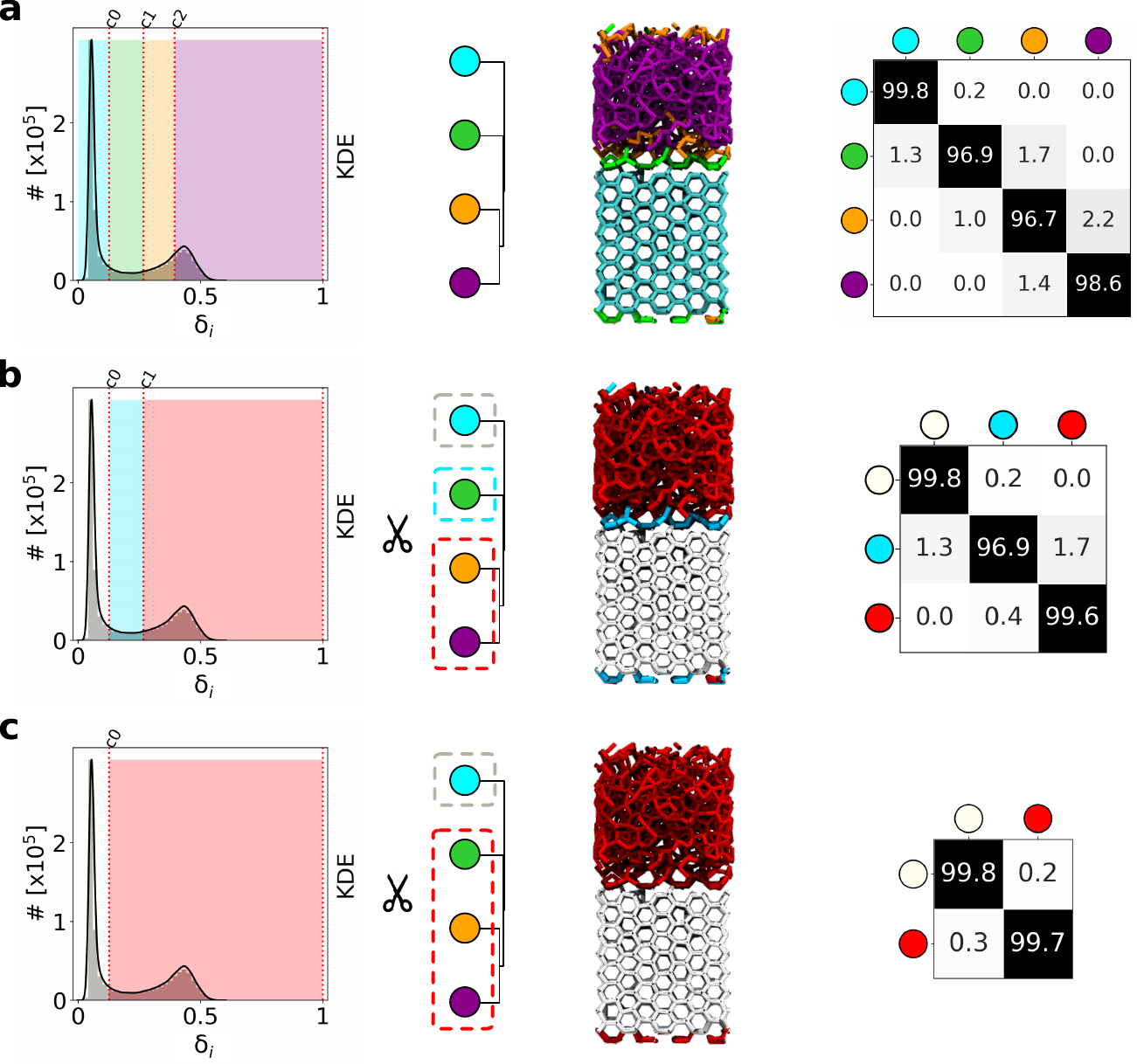}
   \centering
   \caption{\textbf{TIP4P/Ice} froze/melted water: merging LENS clusters at different levels (no merging (a), two clusters merged into one (b) and three clusters merged into one (c)) following the hierarchy given by the dendrogram, example snapshot and transition probability matrix.}
   \label{SIfig07}
\end{figure}

\begin{figure}
 \includegraphics[width=\columnwidth]{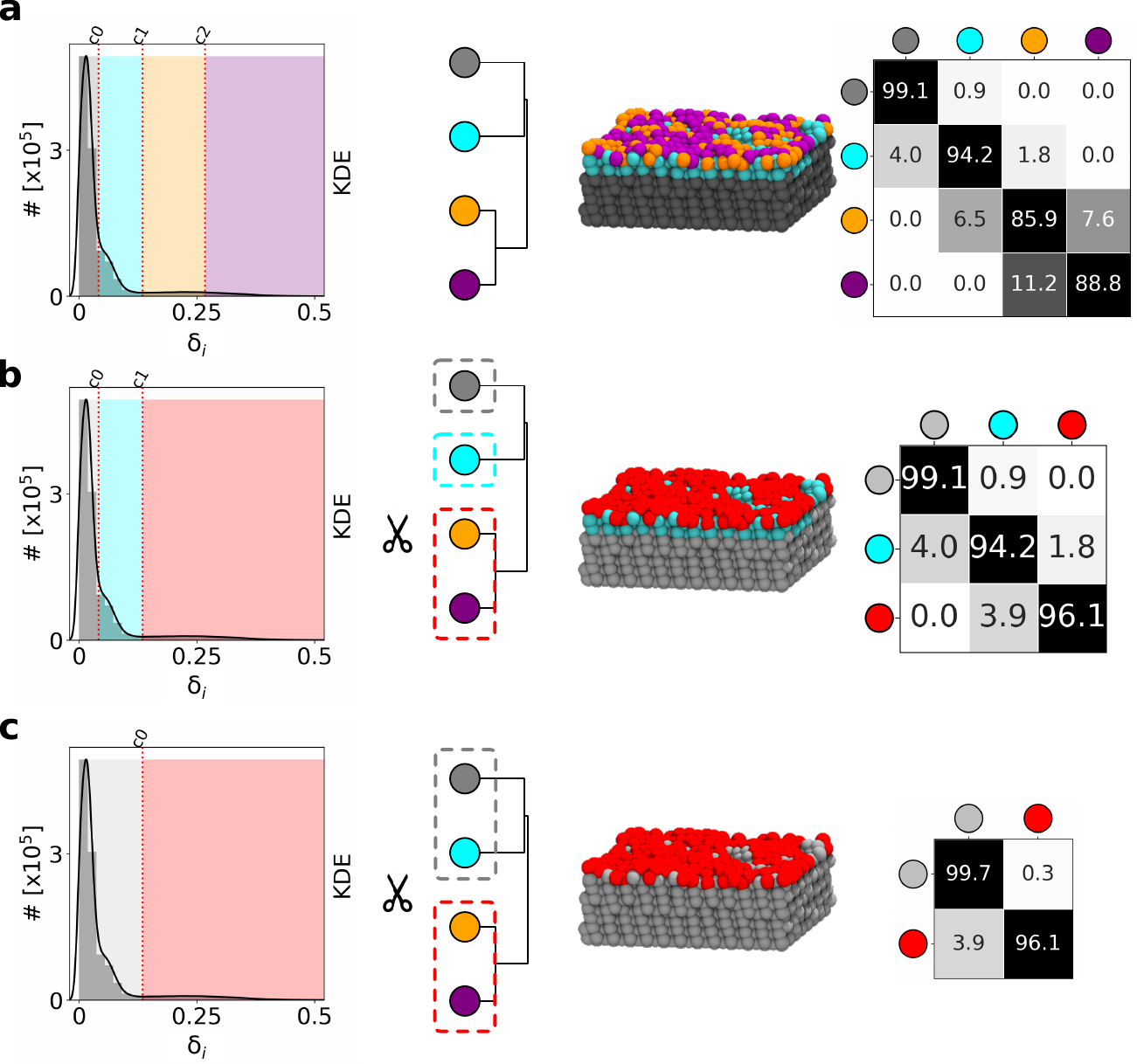}
   \centering
   \caption{\textbf{Cu (210)} copper slab at $T=700$ \si{\kelvin}: merging LENS clusters at different levels (no merging (a), two clusters merged into one (b) and four clusters merged into two (c)) following the hierarchy given by the dendrogram, example snapshot and transition probability matrix.}
   \label{SIfig08}
\end{figure}

\begin{figure}
 \includegraphics[width=\columnwidth]{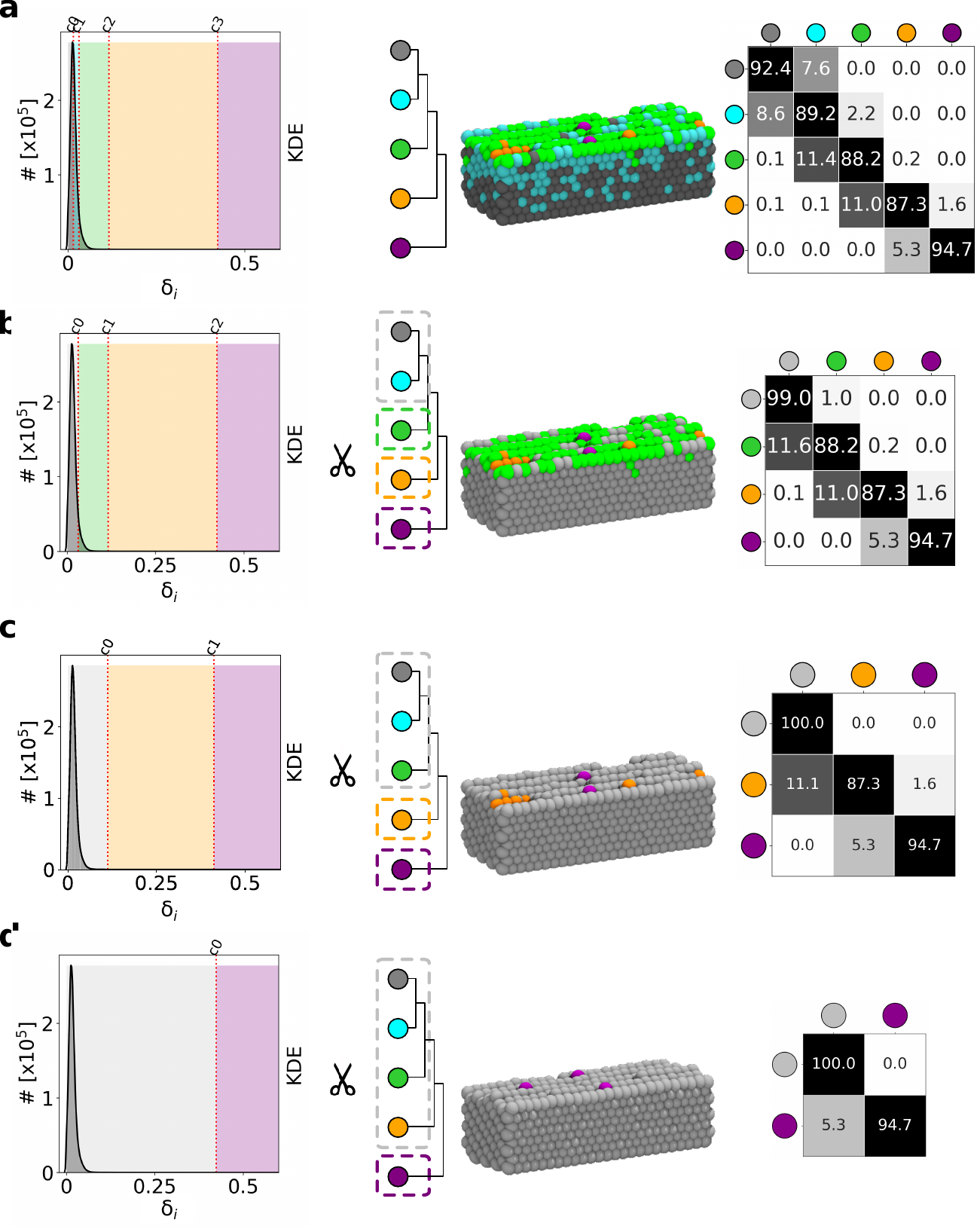}
   \centering
   \caption{\textbf{Cu (211)} copper slab at $T=600$ \si{\kelvin}: merging LENS clusters at different levels (no merging (a), two clusters merged into one (b), three clusters merged into one (c) and four clusters merged into one (d)) following the hierarchy given by the dendrogram, example snapshot and transition probability matrix.}
   \label{SIfig09}
\end{figure}

\begin{figure}
 \includegraphics[width=\columnwidth]{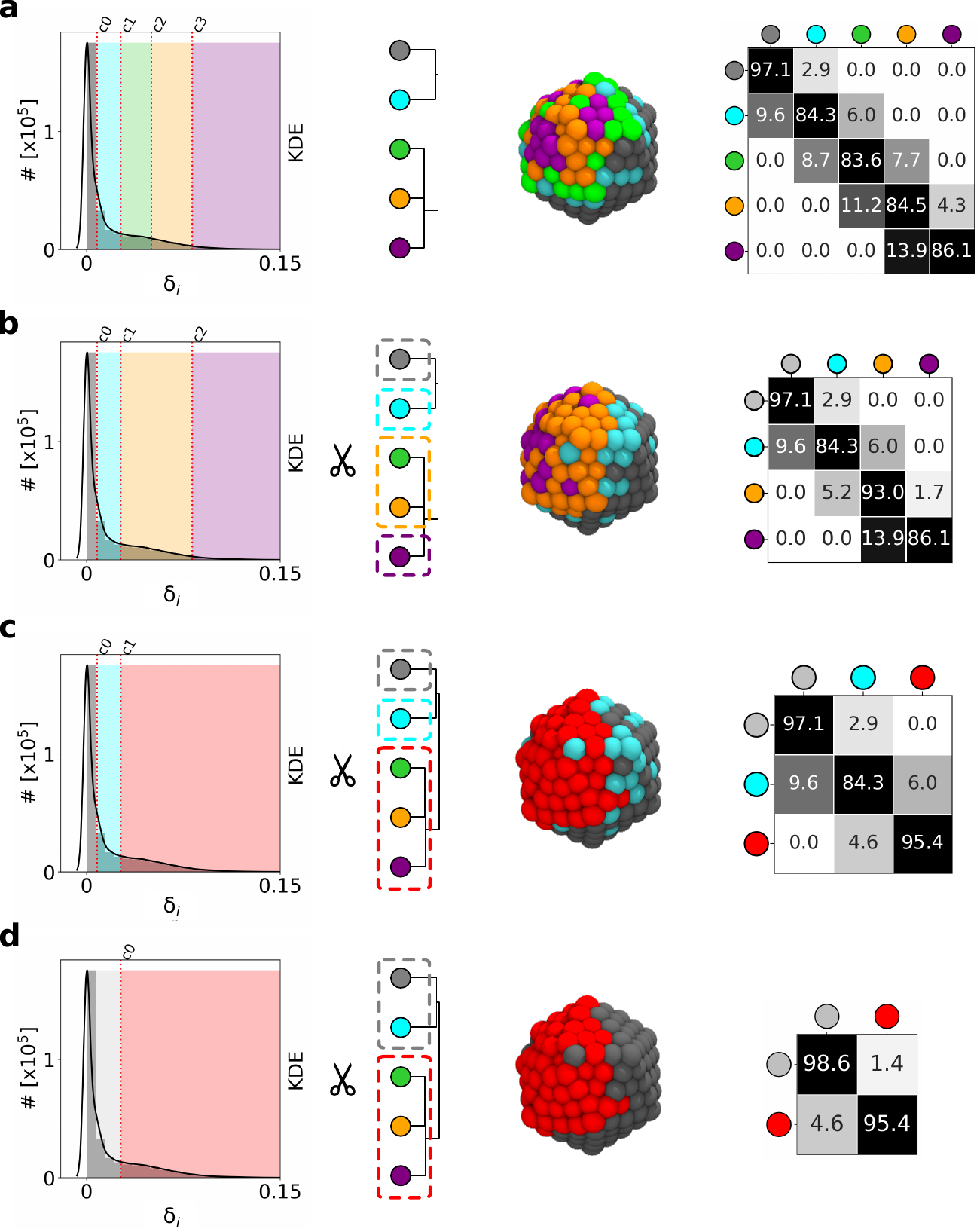}
   \centering
   \caption{\textbf{Au-NP} nanoparticle $T=200$ \si{\kelvin}: merging LENS clusters at different levels (no merging (a), two clusters merged into one (b), three clusters merged into one (c) and three clusters and two clusters merged into two (d)) following the hierarchy given by the dendrogram, example snapshot and transition probability matrix.}
   \label{SIfig10}
\end{figure}

\begin{figure}
 \includegraphics[width=\columnwidth]{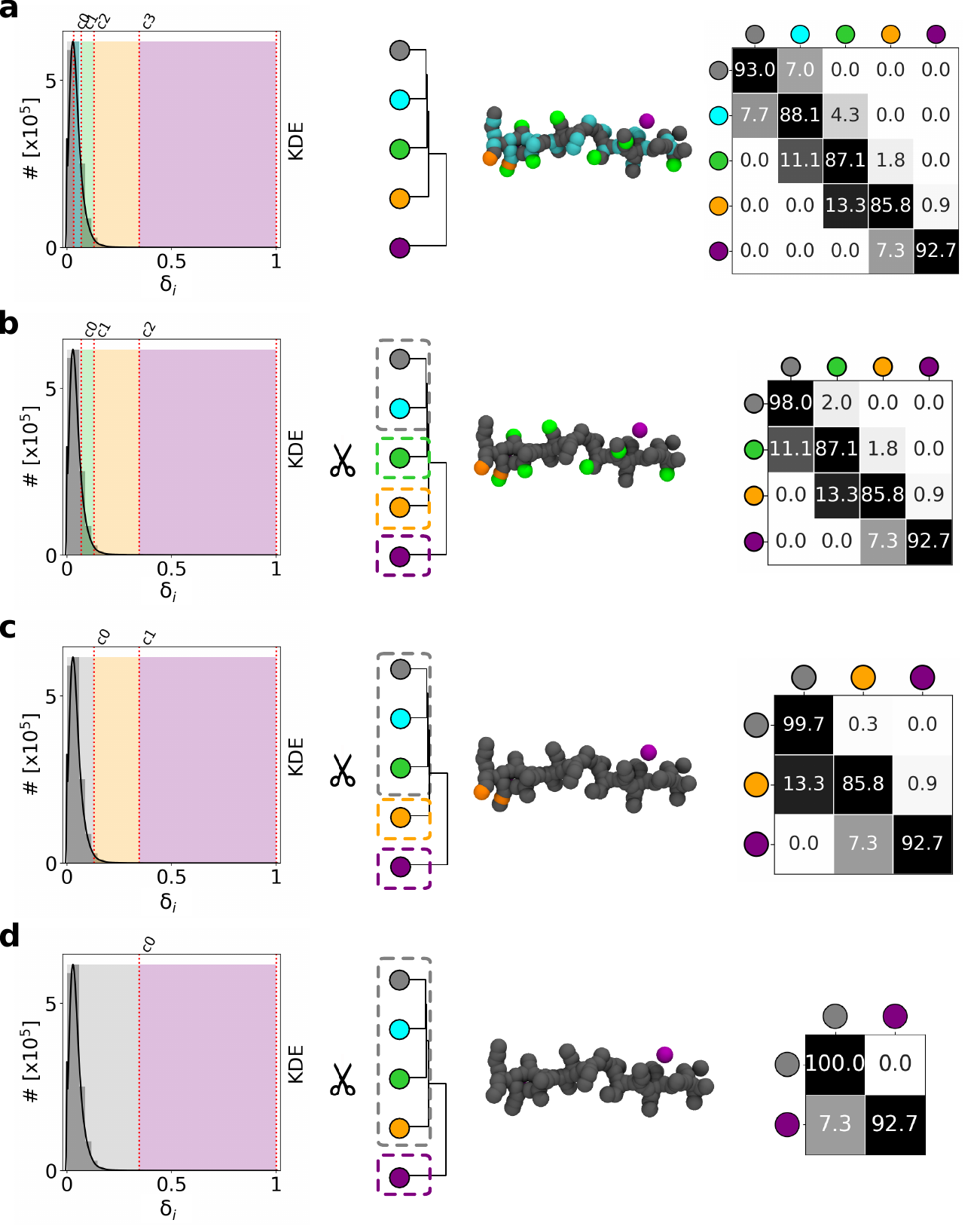}
   \centering
   \caption{\textbf{BTA} fiber: merging LENS clusters at different levels (no merging (a), two clusters merged into one (b), three clusters merged into one (c) and four clusters merged into one (d)) following the hierarchy given by the dendrogram, example snapshot and transition probability matrix.}
   \label{SIfig011}
\end{figure}

\begin{figure}
 \includegraphics[width=\columnwidth]{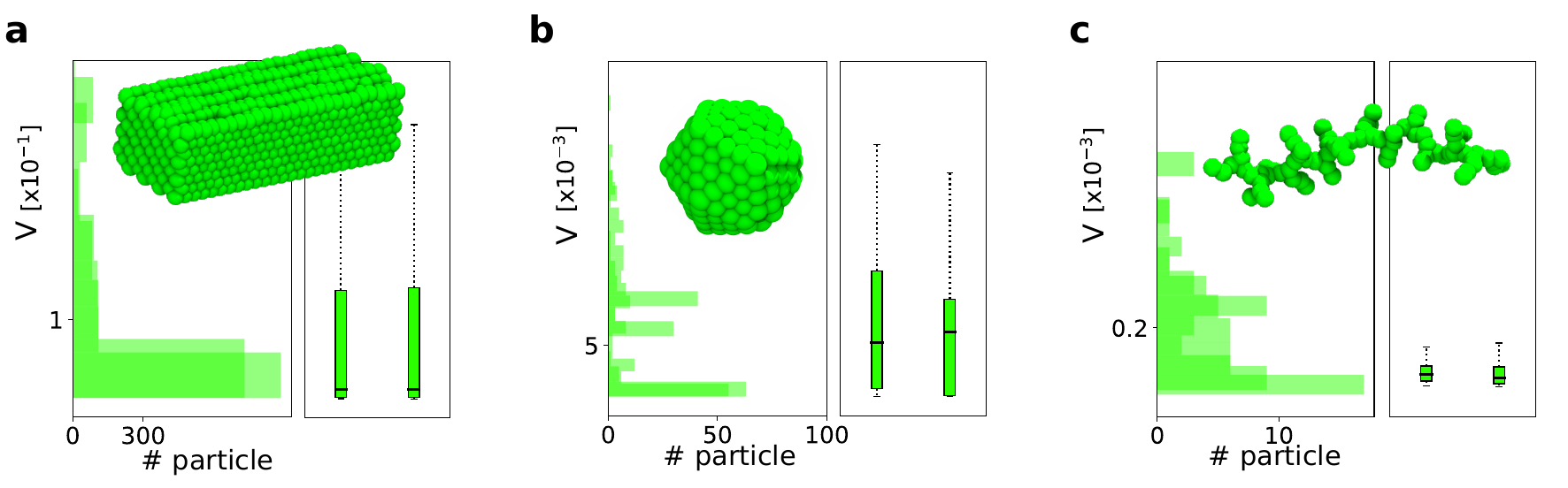}
   \centering
   \caption{Time-independent statistical analysis for \textbf{Cu (211)} copper slab at $T=600$ \si{\kelvin}(a), \textbf{Au-NP} nanoparticle at $T=200$ \si{\kelvin}(b) and \textbf{BTA} fiber (c): when the system is characterized by discrete and fluctuation-like dynamics, a time-indepentend averaged analysis fails to recognize patterns which are not statistically relevant. For example, in the systems reported above, the two clusters identified by HC have identical variability $V$ and they can be classified as the same cluster.}
   \label{SIfig012}
\end{figure}